\documentclass[twocolumn]{aastex63}
\usepackage{amsmath}
\usepackage{cool}
\usepackage{booktabs}
\usepackage{lipsum}
\usepackage{enumitem}

\received{July 10, 2020}
\revised{August 31, 2020}
\accepted{September 7, 2020}
\submitjournal{ApJ}

\shorttitle{Modeling CN Zeeman Observations in Disks}
\shortauthors{Mazzei et al.}

\begin{document}

\title{Untangling Magnetic Complexity in Protoplanetary Disks with the Zeeman Effect}

\correspondingauthor{Renato Mazzei}
\email{mazzei@virginia.edu}

\author[0000-0003-2668-1054]{Renato Mazzei}
\affiliation{Department of Astronomy, University of Virginia, Charlottesville VA 22904, USA}

\author[0000-0003-2076-8001]{L. Ilsedore Cleeves}
\affiliation{Department of Astronomy, University of Virginia, Charlottesville VA 22904, USA}

\author{Zhi-Yun Li}
\affiliation{Department of Astronomy, University of Virginia, Charlottesville VA 22904, USA}

\begin{abstract}
With the recent advent of circular polarization capabilities at the Atacama Large Millimeter/submillimeter Array (ALMA), Zeeman effect measurements of spectral lines are now possible as a means to directly probe line-of-sight magnetic fields in protoplanetary disks (PPDs).
We present a modeling study that aims to guide physical interpretation of these anticipated observations.
Using a fiducial density structure based on a typical ringed disk, we simulate line emission for the hyperfine components of the CN $J = 1-0$ transition with the POLARIS radiative transfer code.
Since the expected magnetic field and typical CN distribution in PPDs remain largely unconstrained, we produce models with several different configurations.
Corresponding integrated Stokes $I$ and $V$ profiles and 0.4 km/s resolution, 1'' beam convolved channel maps are presented.
We demonstrate that the emission signatures from toroidally dominated magnetic fields are distinguishable from vertically dominated magnetic field based on channel map morphology.
Due to line-of-sight and beam cancellation effects, disks with toroidal $\boldsymbol{B}$-field configurations result in significantly diminished Stokes $V$ emission.
Complex magnetic fields therefore render the traditionally used method for inferring line-of-sight magnetic field strengths (i.e., fitting the derivative of the Stokes $I$ to the Stokes $V$ profile) ambiguous, since a given intrinsic field strength can yield a variety of Stokes $V$ amplitudes depending on the magnetic field geometry.
In addition, gas gaps can create structure in the integrated Stokes $V$ profile that might mimic magnetic substructure.
This method should therefore be applied with caution in PPD environments, and can only confidently be used as a measure of magnetic field strength if the disk's magnetic field configuration is well understood.

\end{abstract}


\section{Introduction} \label{sec:intro}
Protoplanetary disks (PPDs) are produced by the gravitational collapse and angular momentum mediated flattening of dense rotating cores in molecular clouds. Their initial formation and subsequent evolution will be strongly impacted by the presence or absence of a magnetic field \cite[e.g.,][and references therein]{li2014}.
Observations of (sub)millimeter continuum dust polarization in cloud complexes reveal suggestive (e.g., ``hour-glass") linear polarization patterns on $\lesssim$1000 AU scales in both low- and high- mass regimes \citep{girart2006,beltran2019}. This structure is commonly interpreted as evidence of magnetic field structure in these environments, with the polarization thought to arise from alignment, through ``radiative torques," of dust grains orthogonal to the local magnetic field \citep{lazarian2007}.
Indeed, a magnetic field with this morphology (pinched toward the center of the collapsing core) is consistent with standard theoretical models for magnetized star formation \citep{galli1993,fielder1993}.

The magnetism of interstellar clouds has also been probed by Zeeman splitting measurements
(e.g., of CN, OH and HI), and studies to this end \citep{falgarone2008,troland2008,heiles2004} reveal that cores are moderately magnetized, with mean line-of-sight $\boldsymbol{B}$-field strengths up to $\approx30$ $\mu$G.
\citet{crutcher2010} concluded through Bayesian analysis of a large sample of dense cores that the most strongly magnetized cores have approximately critical mass-to-flux ratios, suggesting a dynamically important magnetic field regulating the star formation infall process.

Since PPDs form in molecular cloud core environments, it would not be surprising if they inherit some seed magnetization as well, which could be amplified by sheering effects within the disk.
It is difficult to determine the magnetic field morphology of a protoplanetary disk based on core-scale constraints, however, because a large amount of physical evolution and dynamical processing occurs as the disk forms \citep{li2014}.
For example, as gas flows onto the proto-stellar disk and local densities increase, the ionization level drops sufficiently low that non-ideal MHD effects, such as ambipolar diffusion, the Hall effect, and Ohmic dissipation, become important \citep[for review, see e.g.,][]{armitage2019}.
Simulation work that incorporates these physics has been successful in informing how PPDs evolve dynamically under these conditions \citep{turner2014}, but there remains significant ambiguity in determining what constitutes a reasonable initial set-up. We do not have firm answers to some basic questions. How strong should the magnetic field be? How should it be configured?

These questions are of critical importance, as $\boldsymbol{B}-$fields remain central to the study of PPDs and are thought to play a key role in
gas dynamics, which in turn controls the concentration and growth of dust grains that are crucial to the formation of planetesimals and eventually planets \citep{armitage2019}.
In particular, magnetic fields can cause magneto-rotational instability \citep[MRI;][]{balbus1991}, which is widely believed to be a
dominant driver of gas accretion in disk systems.
This interpretation remains uncertain in light of observations that suggest
ionization rates  that are too low for the MRI to operate efficiently \citep{cleeves2015}, which is consistent with the low levels of turbulence inferred in some disks \citep[e.g.,][]{flaherty2015}.
Poloidal field components may also launch jets and winds perpendicular to the
disk plane \citep[e.g.,][]{blandford1982, simon2013} that
mediate gas accretion.
These flows have been proposed to trigger the formation of rings and gaps \citep{suriano2017}, and field-dependent mechanisms \citep[e.g., ``zonal flows,"][]{johansen2009,bai2013} can lead to planetesimal formation as well.

Since there is a wealth of disk physics that depends on the magnetic field strength and orientation, observational constraints are important.
To date there has never been an independently confirmed direct measurement of a magnetic field in a protoplanetary disk.
This is largely because linear polarization, the historically available technique for inferring magnetic information, has yielded results on the disk-scale that are difficult to reconcile with any clear $\boldsymbol{B}-$field interpretation.
Though magnetic alignment is expected \citep{cho2007,bertrang2017},
recent work has demonstrated that a variety of other mechanisms may also produce millimeter linear polarization in disks, including self-scattering of thermal dust emission \citep{kataoka2015,yang2016}, radiation field (e..g ``k-RAT") alignment \citep{kataoka2017,tazaki2017}, and gas flow alignment \citep{kataoka2019}, none of which depend explicitly on the magnetic field geometry.

Fortunately, circular polarization is now possible with the Atacama Large Millimeter/submillimeter Array (ALMA), providing us with the opportunity to carry out ``Zeeman effect" observations as a more definitive technique for accessing line-of-sight magnetic information in PPDs \citep[e.g.,][]{vlemmings2019}.
With more observations on the horizon, this paper aims to elucidate physical interpretation of disk-scale circular polarization and address the main difficulties associated with inferences of magnetic structure in PPDs.
We perform full radiative transfer simulations of Zeeman observations of the CN $J = 1 - 0$ transition for several different disk set-ups (in terms of CN distribution and magnetic field configuration), then interpret the emission and assess its detectability under a variety of conditions.
Finally, we address the importance of beam size, which presents challenges that are unique to circular polarization observations.

\section{Zeeman Effect Primer} \label{sec:zeemanreview}
\begin{figure*}
  \includegraphics[width=1.05\textwidth]{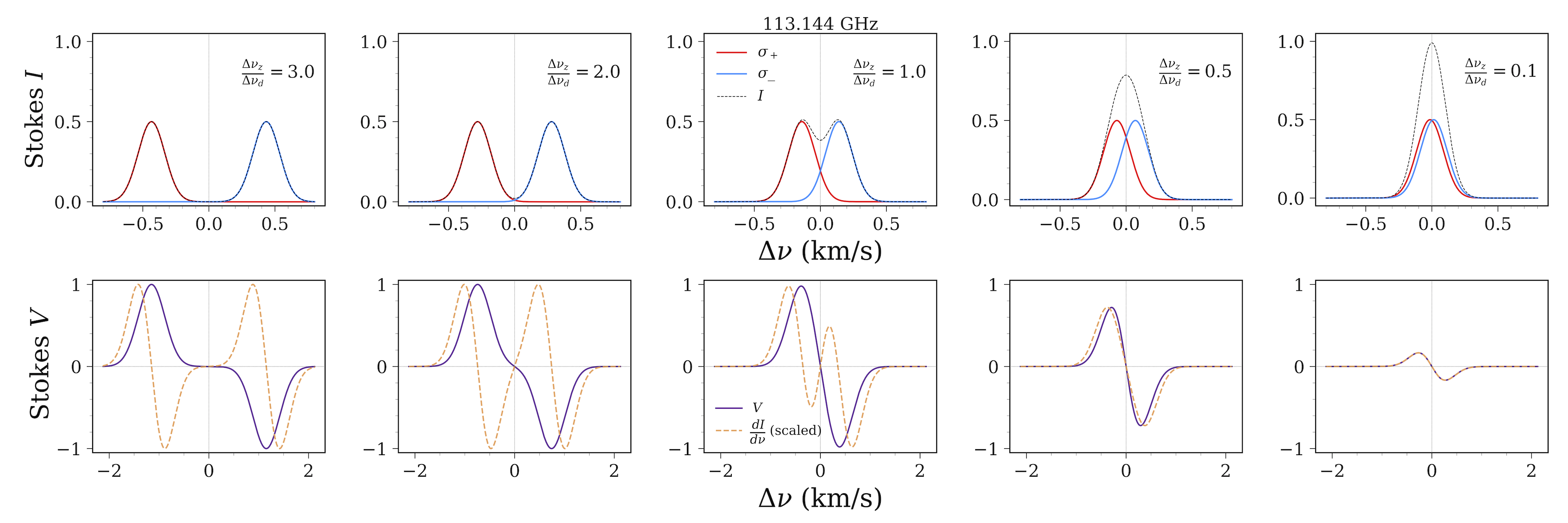}
  \caption{Stokes $I$ (top) and Stokes $V$ (bottom) profiles for a variety of choices of $\Delta \nu_z / \Delta \nu_{\rm line}$ in the case of Doppler (i.e., Gaussian) broadened lines. In this demonstration we set $T = 20$ K, $\nu_0 = 113.144$ GHz (the frequency of the CN $J = 1 - 0$ transiton), and calculate $\Delta \nu_d = \frac{\nu_0}{c}\sqrt{\frac{2kT}{m_p}}$. We then vary $B$ to calculate each $\Delta \nu_z$. In the unresolved limit, the magnitude of $V$ scales linearly with the magnetic field strength. Protoplanetary disks fall in this regime, since their field strengths are expected to be relatively weak. Note that in each plot on the bottom row, the $\frac{dI}{d\nu}$ curve has been scaled down to match the magnitude of the $V$ curve for figure clarity.}
  \label{fig:splitDemo}
\end{figure*}

For a parcel of gas threaded by a magnetic field, Zeeman-sensitive species' line emission is split into two circularly polarized components:
\begin{itemize}[topsep=8pt,itemsep=4pt,partopsep=4pt, parsep=4pt]
    \item[] $\sigma_+(\nu)$: line center at $\nu = \nu_0 - \Delta \nu_z$,
    \item[] $\sigma_-(\nu)$: line center at $\nu = \nu_0 + \Delta \nu_z$
\end{itemize}
where $\Delta \nu_z = z_B B$.
The value of $z_B$ (the so-called ``Zeeman-factor") is calculated as
\begin{equation}
    z_B = \frac{\mu_B}{h}(g'M'-g''M'')\,,
\end{equation}
where $\mu_B$ is the Bohr magneton, $g$ is the Land\'e factor, and $M$ is the projection of the total angular momentum quantum number onto the magnetic field.
Single and double primes represent the upper and lower levels of a given line transition, respectively.
The $\sigma_+$ and $\sigma_-$ line profiles have the same intrinsic width, $\Delta \nu_{\rm line}$, as determined by the typical environmental processes (e.g., thermal, pressure, natural broadening),
and the Stokes $I$ and Stokes $V$\footnote{The choice $V = \sigma_+ - \sigma_-$ (instead of $V = \sigma_- - \sigma_+$) is purely a matter of convention.} of the emission are as follows:
\begin{align}
    I(\nu) &= \sigma_+(\nu) + \sigma_-(\nu) \\
    V(\nu) &= \sigma_+(\nu) - \sigma_-(\nu) \,.
\end{align}
If the magnetic field is uniform along the line-of-sight and sufficiently weak such that $\Delta \nu_z \lesssim \Delta \nu_{\rm line}$ (i.e., unresolved splitting, which is expected for both molecular cloud-like and PPD environments), the Stokes $V$ can be related approximately to the Stokes $I$ as
\begin{align}
    V = \frac{dI}{d\nu} \Delta \nu_z \cos{\theta_B}\,. \label{eq:VB}
\end{align}
Here, $\theta_B$ is the inclination of the magnetic field relative to the line-of-sight \citep{crutcher1993}.
In Figure \ref{fig:splitDemo}, we demonstrate the $I$ and $V$ profiles obtained from Doppler broadened lines for a variety of choices of $\Delta \nu_z / \Delta \nu_{\rm line}$, showing that Equation \ref{eq:VB} is an increasingly good approximation in the $\Delta \nu_z \rightarrow 0$ limit.
It is worth stressing that $dI/d\nu$ mimicks the shape of the $V$ profile only for \textit{uniform} magnetic fields.
Magnetic field configurations with significant sub-structure (e.g., toroidal or radial components) can cause the relationship to break down.

\section{Parametric Modeling}\label{sec:model}
We explore a fully parametric disk model for use in our radiative transfer simulations, to allow us to probe a variety of disk chemical and physical set-ups.
We first produce ``simple'' models with purely vertical and purely toroidal field configurations as morphological case studies.
We then invoke a more complex magnetic field geometry and vary parameters related to the distribution of our emitting molecule (CN) and the magnetic field strength.
A description and list of chosen values for our parameter exploration is given in Table \ref{tab:parVary}.
Our fiducial disk structure is inspired by AS 209, a nearby ($d \approx 126$ pc), approximately solar-mass star with a minimally extincted \citep[$A_{\rm v}$ = 0.8;][]{avenhaus2018}, moderately inclined ($i = 38^{\circ}$) disk that has been observed to have CN $J = 2-1$ emission to a radial extent of $\sim 200$ AU \citep{oberg2011}.
These favorable observational characteristics have made AS 209 a common choice for pilot circular polarization studies with ALMA (e.g., 2018.1.01030.S, PI: R. Harrision; 2018.1.00298.S, PI: L. Cleeves).
It should be noted that though we use gas and dust distributions specifically fitted to AS 209 (see Section \ref{ssec:phys}), the bulk structure is not dissimilar from a variety of other disks \citep{andrews2009}.
In addition, recent sub-millimeter observations \citep[e.g., DSHARP;][]{andrews2018} also show that dust sub-structure is common in PPDs.
Therefore, the model presented in this work is intended to serve as an example of a ``typical" disk, and we expect the general trends found here to be broadly applicable.

\begin{deluxetable*}{llll}
\tablecaption{Selected values for our parameter exploration. For each parameter we run a batch of line emission simulations of the 113.144 GHz CN $J = 1-0$ transition over the specified range, with all other parameters set to their fiducial values. \label{tab:parVary}}
\setlength{\tabcolsep}{5pt} 
\renewcommand{\arraystretch}{1} 
\tablehead{Parameter                          & Fiducial Value & Range                          & Description}
\startdata
$\rm X_{\rm CN}$                  & $\mathbf{10^{-8}}$ &  $5 \times 10^{-10}$ - $5 \times 10^{-7}$  & CN abundance in slab (relative to $H_2$)                                                            \\
$R_{\rm in, CN}$ (AU)                  & \textbf{30}    & 1 - 60                   & Inner radius of CN slab                                                          \\
$R_{\rm out, CN}$ (AU)                 & \textbf{150} & 90 - 200                & Outer radius of CN slab                                                          \\
$N_{\rm min, CN}$ ($\times 10^{21}$ cm$^{-2}$) & \textbf{0.5} & 0.05 - 3                            & Minimum column density of CN slab                                                \\
$N_{\rm max, CN}$ ($\times 10^{21}$ cm$^{-2}$) & \textbf{10} & 5 - 200                     & Maximum column density of CN slab                                                \\
$B_{\rm sum,0}$ (mG)               & \textbf{40} & 5 - 100                     & Sum\tablenotemark{a} of magnetic field components at $R = 1$ AU                                            \\
$\beta_{B_{\rm r}}$ & \textbf{-0.75} & -0.3 to -1.3&  Power law index for radial falloff in magnetic field strength \\
$f_1$                     & \textbf{0.3}                      & - & $B_{\rm vert,0}$/$B_{\rm sum,0}$                        \\
$f_2$                      & \textbf{0.36}                      & - & $B_{\rm rad,0}$/($B_{\rm sum,0} (1-f_1)$)  \\
$i$ ($^{\circ}$)                             &  - & 0, 40, 90                    & Disk inclination ($0^{\circ} =$ face-on, $90^{\circ} =$ edge-on)                                   \\
$f_{\rm lg}$ & \textbf{0.85} & - &  Fraction of $M_{\rm dust}$ put into the large dust population \\
\enddata
\tablenotetext{a}{$B_{\rm sum,0}$ = $B_{\rm rad,0}$ + $B_{\rm tor,0}$ + $B_{\rm vert,0}$, the sum of the radial, toroidal, and vertical magnetic field components, respectively.}
\end{deluxetable*}

\subsection{Density Structure}\label{ssec:phys}
Our gas density distribution is based on the best-fit self-similar accretion disk solution obtained through multi-wavelength fitting of AS 209 by \citet{tazzari2016}.
Their reported gas surface density profile has a power-law falloff and exponential taper
\begin{align}
    \Sigma_{\rm g}(R) = \Sigma^0_{\rm g} \big(\frac{R}{R_0}\big)^{\gamma_0} e^{-\big(\frac{R}{R_{\rm c}}\big)^{2+\gamma_0}}
\end{align}
with parameter choices $R_0 = 40$ AU, critical radius $R_{\rm c} = 78$ AU, and $\gamma_0 = -0.91$.

Dust plays an important role in radiative transfer and should be modeled as accurately as possible to produce a reasonable calculation  of the disk's temperature.
We include two dust density distributions to simultanously account for a puffed-up, hydrostatically supported layer of small grains and a midplane-settled population of large grains.
Both are set to have MRN \citep{mathis1977} power-law size distributions, with the small population ranging from 0.005-1 $\mu$m and the large population ranging from 0.005-2000 $\mu$m.
We take the small dust to be spatially co-located with the gas, and set the large dust distribution based on the best-fit surface density profile from ALMA 1.3 mm observations \citep{fedele2018},
\begin{align}
    \Sigma_{\rm d,lg}(R) = \Sigma^0_{\rm d,lg} \delta(R) \big(\frac{R}{R_{\rm c}}\big)^{\gamma_1} e^{-\big(\frac{R}{R_{\rm c}}\big)^{\gamma_2}} \,,
\end{align}
with $\gamma_1 = 0.3$ and $\gamma_2 = 2.0$. The scaling parameter $\delta(R)$ models the observed ring/gap sub-structure in AS 209 and is written as
\begin{align}
\delta(R)=
\begin{cases}
\delta_{\rm G1} = 0.1         & \text{$R$ $\in$ [$R_{\rm G1} - hw_{\rm G1}, R_{\rm G1} + hw_{\rm G1}$]}\\
\delta_{\rm R1} = 0.75         & \text{$R$ $\in$ [$R_{\rm G1} + hw_{\rm G1}, R_{\rm G2} - hw_{\rm G2}$]}\\
\delta_{\rm G2} = 0.01         & \text{$R$ $\in$ [$R_{\rm G2} - hw_{\rm G2}, R_{\rm G2} + hw_{\rm G2}$]}\\
\delta_{\rm R2} = 4.5         & \text{$R$ $\in$ [$R_{\rm G2} + hw_{\rm G2}, R_{\rm R2, out} $] } \\
\delta_{\rm out} = 1.5        & \text{$R$ $\geq  R_{\rm R2, out}$}\\
1                                & \text{otherwise}\,,
\end{cases} \label{eq:delta}
\end{align}
with the gaps parameterized by best-fit radii ($R_{\rm G1} = 62$ AU, $R_{\rm G2} = 103$ AU)  and half-widths ($hw_{\rm G1} = 8$ AU, $hw_{\rm G2} = 16$ AU). The outer ring has an outer radius of $R_{\rm R2,out} = 140$ AU. Outside of disk radius $R_{\rm out} = 200$ AU, we set both the gas and dust surface densities to zero.
Our fiducual disk model does not include gas deficits.
However, there is observational evidence from near-infrared scattered light \citep{avenhaus2018} and CO line transition data \citep{favre2019} that gas gaps may be present in AS 209 and similar disks, perhaps co-located with the dust gaps.
We explore their impact on Zeeman observations in Section \ref{ssec:gas}.

\begin{figure*}
  \includegraphics[width=\textwidth]{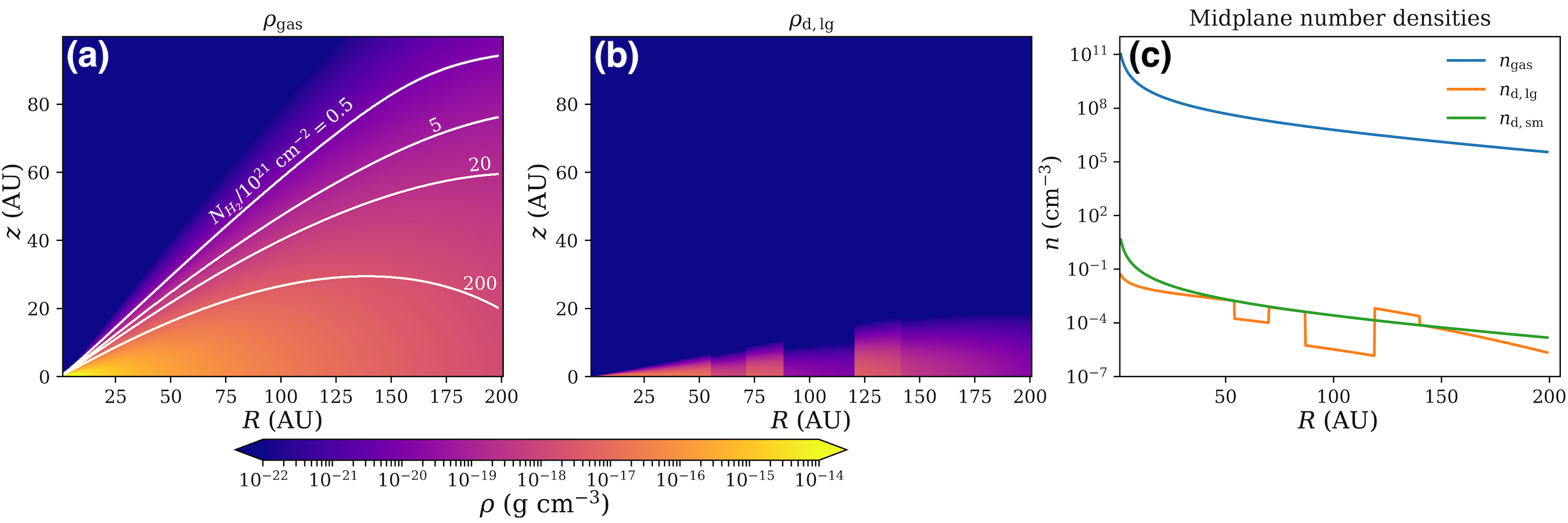}
  \caption{Density structure of our AS 209 based model. \textbf{Panel (a)}: Gas density, overlaid with vertically computed H$_2$ column density contours relative to ($N_{H_2}/10^{21}\text{ cm}^{-2}$).
  $N_{H_2}$ bounds the allowed limits for CN to reside in our simulations. The small dust is co-located with the gas.
  \textbf{Panel (b)}: A large dust density slice. The gaps at 62 AU and 103 AU are based on previous modeling of AS 209's sub-mm dust continuum observations and aim to make our model more realistic due to the observed prevalence of substructure in disks \citep{andrews2018}. \textbf{Panel (c)}: Midplane number densities as a function of radius for our gas, large dust, and small dust distributions.}
  \label{fig:densities}
\end{figure*}

The 2.5-dimensional distributions used in our simulations are constructed from the above detailed 1-dimensional surface density profiles using the general conversion
\begin{align}
    \rho_i(R,z) = \Sigma_i(R) \frac{e^{-\frac{1}{2}(\frac{\theta_z}{h_i})^2}}{\sqrt{2\pi}R h_i}\,,
\end{align}
where $\theta_z = \arctan{(|z|/R)}$.
The scale height $h_i$ for each distribution allows for flaring and is  parameterized as
\begin{align}
    h_i = \chi_i h_{\rm c} \big(\frac{R}{R_{\rm c}}\big)^{\psi}\,,
\end{align}
where $h_{\rm c}$ is a dimensionless critical scale height (normalized to radius), $R_{\rm c}$ is the critical radius of the disk, and $\psi$ is the disk flaring parameter. For consistency with the \citet{fedele2018} results, we choose $\psi = 0.1$, $h_{\rm c} = 0.133$, $\chi_{\rm g} = \chi_{\rm d,sm} = 1$, and $\chi_{\rm d,lg} = 0.2$, where subscripts correspond to gas, small dust, and large dust, respectively.
We also set the total disk dust mass as $M_{dust} = 3.5 \times 10^{-4}$ $M_{\odot}$. To determine the normalizations for our density distributions, we assume a gas-to-dust mass ratio of 100 and set the fraction of dust mass in the large grain distribution by parameter $f_{\rm lg}$. Numerical integration then easily yields appropriate values for $\Sigma^0_{\rm g}$, $\Sigma^0_{\rm d,sm}$, and $\Sigma^0_{\rm d,lg}$. In Figure \ref{fig:densities} we show edge-on midplane cuts of $\rho_{\rm g}$, $\rho_{\rm d,sm}$, and $\rho_{\rm d,lg}$ for our ``AS 209"-like density structure.

\subsection{Velocity Field}
The bulk gas motions are assumed to be Keplerian, i.e.,
\begin{align}
    \boldsymbol{v}(R,z) = \sqrt{\frac{GM_*}{R}} \boldsymbol{\hat{\phi}}\,,
\end{align}
where $\boldsymbol{\hat{\phi}}$ is the azimuthal unit vector in cylindrical coordinates and $M_*=0.9$ $M_{\odot}$ \citep{andrews2009}.
In addition, the line emission simulations include thermal broadening and an additional turbulent component specified by the user, where we choose $v_{\rm turb} = 0.1$ km/s \citep{pietu2007,chapillon2012}.

\subsection{Magnetic Field}
We adopt a parametric description of the disk magnetic field. The magnetic field strength is set to obey a radial power-law
\begin{align}
    B_{\rm sum}(R) = B_{\rm sum,0} \big(\frac{R}{\rm 1\,  AU}\big)^{\beta_{B_{\rm r}}}\,,
\end{align}
with $B_{\rm sum}$ assumed to be constant as a function of $z$, approximately consistent with the results of magnetized simulations  from \citet{suriano2017} that include a disk-wind. A routinely used method for deriving reasonable values for $\beta_{B_{\rm r}}$ is to invoke self-similarity between the radial gas density and magnetic field strength profiles. Taking $P \propto \rho^{\Gamma}$, it
is straightforward to show \citep{zanni2007} that $\beta_{B_{\rm r}}$ is a function of the radial gas density power law, $\beta_{\rho}$:
\begin{align}
    \beta_{B_{\rm r}} = \frac{\Gamma \beta_{\rho}}{2} \,.
\end{align}
Adopting $\Gamma = 5/3$ and setting $\beta_{\rho} = \gamma_0 = -0.91$ from the AS 209 gas density distribution modeled in Section \ref{ssec:phys}, we obtain $\beta_{B_{\rm r}} = -0.758$.
We use this calculation as a guide for our fiducial value.

At each radial location in the disk, we divide the magnetic field strength into independent toroidal, radial, and vertical components as
\begin{multline}
    \boldsymbol{B}(R,z) = (1 - f_1) f_2 B_{\rm sum} \hat{\boldsymbol{r}}^{\prime}+ \\ (1 - f_1) (1-f_2) B_{\rm sum} \hat{\boldsymbol{\phi}}^{\prime} + f_1 B_{\rm sum} \hat{\boldsymbol{z}}\,,
\end{multline}
where $f_1 \leq 1$ and $f_2 \leq 1$.
Also, we prescribe
\begin{align}
  \hat{\boldsymbol{r}}^{\prime} =
  \begin{cases}
    \hat{\boldsymbol{r}} & \text{if } z > 0\\
    -\hat{\boldsymbol{r}}  & \text{otherwise} \,,
  \end{cases} \label{eq:cond1}
\end{align}
and
\begin{align}
  \hat{\boldsymbol{\phi}}^{\prime} =
  \begin{cases}
    \hat{\boldsymbol{\phi}} & \text{if } z > 0\\
    -\hat{\boldsymbol{\phi}}  & \text{otherwise} \,.
  \end{cases} \label{eq:cond2}
\end{align}
Equation \ref{eq:cond1} is included to model the ``wind-up" that occurs in the toroidal $\boldsymbol{B}$-field component due to disk rotation \citep[per simulations, e.g.][]{romanova2012}, and Equation \ref{eq:cond2} accounts for the reversal of the radial component that occurs due to inward dragging in accretion disks.

Our fiducial choices (see Table \ref{tab:parVary}) for $f_1$, $f_2$, and $B_{\rm sum,0}$ are guided by the results of disk wind simulations \citep{suriano2017} after 1000 orbits.
A few other values are also explored to examine a diverse variety of potential magnetic field configurations.

\subsection{CN Distribution}
Chemical modeling of PPDs with many different physical structures by \citet{cazzoletti2018} suggests that it is ubiquitous for CN to reside in a relatively thin layer in the upper and outer regions of the disk.
This structure arises because CN abundance is mainly governed by the balance between ionizing far ultraviolet photons (which produce overwhelming photodissociation and photoionization at $N_{\text{H}_2} \lesssim 10^{20}$ cm$^{-2}$) and freeze-out onto grains deep in the disk at low temperatures, $\lesssim32$~K.
Chemical models also find CN abundances are approximately constant (to within a factor of $\approx$2) within this intermediate layer irrespective of radius, modulo an inner deficit of CN.
Given these constraints, we set the distribution of CN in our simulations to be a constant abundance slab.
The slab is defined to have inner and outer radii, $R_{\rm in,CN}$ and $R_{\rm out,CN}$, and the vertical extent is set by upper and lower H$_2$ column densities, $N_{\rm min,CN}$ and $N_{\rm max,CN}$.
Expected values for $N_{\rm in,CN}$, $N_{\rm out,CN}$, $R_{\rm in,CN}$, and $R_{\rm out,CN}$ are not precisely constrained, so we vary each over a few different reasonable possibilities in Section \ref{sssec::parsli}.

\section{Simulation Methods}\label{sec:simmeth}
We perform our simulations using the POLARIS 3D radiative transfer code \citep{reissl2016,brauer2017b}.
Radiative transfer in POLARIS is solved using Mol3D \citep{ober2015}, and spectral line Zeeman splitting and polarization is based on the Stokes formalism implementation by \citet{larsson2014}.
We specify physical quantities in an octree format,
with grid sub-division set using a variable refinement scheme based on local gas density.
The densest regions have $\sim 0.2$ AU resolution, with reduced resolution approximately linearly down to $\sim 8$ AU in the most diffuse parts of the disk, such as the upper atmosphere above the CN emitting region.
Each simulation involves two computations: first a temperature calculation based on the dust density structure, then the CN line emission. Each step is detailed further in the following sections.

\subsection{Temperature Calculation}\label{ssec:rt}
The disk is heated by irradiation from a central point source, set to have luminosity consistent with a blackbody that has AS 209 stellar parameters \citep[$R = 2.3$  R$_{\odot}$, $T = 4250$~K; ][]{tazzari2016}.
We use $10^7$ photons in this calculation to ensure good coverage in all regions of the disk.
After each photon is generated (with characteristic wavelength, energy per unit time, and randomly chosen direction), it is allowed to scatter on dust grains according to an isotropic phase function. Dust heating is handled with continuous absorption \citep{lucy1999} and immediate re-emission \citep{bjorkman2001} methods.
After all photons from the central star have been propagated, $T_{\rm dust}$ at each location in the disk is determined based on the temperature of local grains.
We then set $T_{\rm gas} = T_{\rm dust}$ for simplicity in our parametric model; however, we note that the disk gas in the atmosphere is likely warmer than the dust temperature, due to additional UV heating from the star. This could result in generally brighter CN emission than what is predicted here.

\subsection{Emission from CN Spectral Lines}\label{ssec:zeeman}
The $J = 1-0$ transition of CN presents nine hyperfine Zeeman components, seven of which are strong enough to be of potential astronomical relevance. In Table \ref{tab:CNLines} we give the rest frequency ($\nu_0$), relative intensity (RI), and Zeeman factor ($z_B$) for each of these lines, as originally tabulated by \citet{falgarone2008}.
For our main set of models we only consider the 113.144 GHz transition, since it is a good representative case with high relative sensitivity to $B_{\rm LOS}$ and a large $z_B$.
In Section \ref{sssec:transitions}, we simulate (and stack) the emission from all seven lines for our fiducial disk.

\begin{deluxetable}{cccc}
\tablecaption{The seven strong hyperfine lines for the CN $J = 1-0$ transition. RI$\times$$z_b$ quantifies relative sensitivity to $B_{\rm LOS}$. \label{tab:CNLines}}
\setlength{\tabcolsep}{5pt} 
\renewcommand{\arraystretch}{1} 
\tablehead{$\nu_0$ (GHz) & RI & $z_B$ (Hz/$\mu$G) & RI$\times z_B$}
\startdata
113.144       & 8  & 2.18              & 17.4              \\
113.171       & 8  & -0.31             & 2.5               \\
113.191       & 10 & 0.62              & 6.2               \\
113.488       & 10 & 2.18              & 21.8              \\
113.491       & 27 & 0.56              & 15.1              \\
113.500       & 8  & 0.62              & 5.0               \\
113.509       & 8  & 1.62              & 13.0              \\
\enddata
\end{deluxetable}

Zeeman-splitting line emission in POLARIS is computed using the ZRAD extension \citep{brauer2017b}.
ZRAD makes use of energy level and transition data from the Leiden Atomic and Molecular DAtabase \citep[LAMDA;][]{schoier2005} and the JPL spectral line catalog \citep{pickett1998}.
This work uses the CN hyperfine data set, with rates from \citet{kalugina2015}.
Natural, collisional, and Doppler broadening, as well as the magneto-optic effect \citep{larsson2014}, are all considered in determining the line shape, and the final profile is calculated with a Faddeeva function solver\footnote{http://ab-initio.mit.edu/wiki/index.php/Faddeeva\_Package, Copyright \copyright 2012 Massachusetts Institute of Technology}.
For the turbulent component we choose $v_{\rm turb} = 0.1$ km/s \citep{pietu2007,chapillon2012}, or about 30\% of the sound speed. 

We initialize our line radiative transfer simulations with $10^5$ unpolarized background photons and assume local thermodynamic equilibrium (LTE) for all level population calculations.
Photons are ray traced to a $256\times 256$ pixel detector, where the Stokes $I$ and $V$ of the emission are recorded.
We set the detector to observe in 181 velocity channels in range $[v_0-6\text{ km/s},v_0+6\text{ km/s}]$, producing 0.067 km~s$^{-1}$ resolution data. The source velocity is set to $v_0 = 0$ km~s$^{-1}$.

\section{Results}\label{sec:results}
Our POLARIS simulations yield 3D data cubes with spatially resolved $I$, $V$, and optical depth ($\tau$) information for each pixel in each of the 181 channels.
We then bin the data to 0.4 km/s wide frequency bins
and convolve the data with a Gaussian kernel to simulate a $1''$ beam.
From these processed data, we produce channel maps and spatially integrated line profiles.

\subsection{Vertical and Toroidal Magnetic Field Case Studies}\label{ssec:magCases}
Presented here are the results of simulations with either vertical or toroidal magnetic field configurations. All the parameters from Table \ref{tab:parVary} (except $f_1$ and $f_2$) are set to their fiducial values for these models, except for the maximum column density of the CN slab which we set to $N_{\rm max, CN} = 20 \times 10^{21}$ cm$^{-2}$ here. While this choice is arbitrary, it ensures that the CN is not too optically thick such that the Stokes $V$ is dominated by magnetic effects rather than opacity.
Opacity varies due to the geometry of the CN emitting gas and sight line effects, but aside from some regions in the vertical magnetic field case when viewed face-on,
$\tau < 1$ at all locations in observer space across all frequencies for these runs.
Therefore, these models are reasonable approximations of the ``optically-thin" limit.

\subsubsection{Vertical Magnetic Field}\label{sssec:vertCase}
\begin{figure*}
  \makebox[\textwidth][c]{\includegraphics[width=0.99\textwidth]{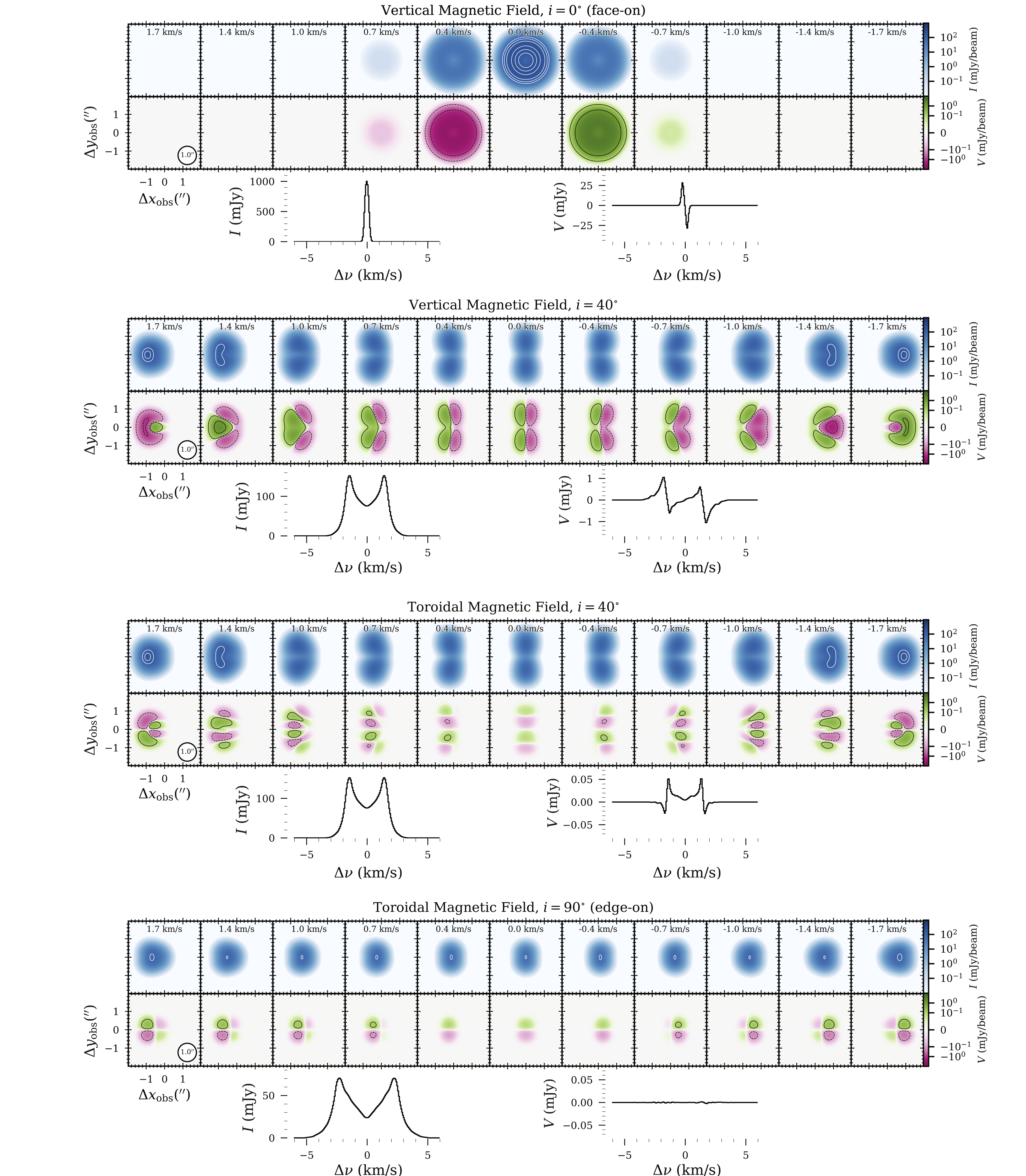}}
  \caption{Model Stokes $I$ and $V$ channel maps of the 113.144 GHz CN $J = 1 - 0$ transition. The top two panels show vertical magnetic field simulations at the labelled inclinations. The bottom two panels show toroidal magnetic field simulations at edge on and $40^{\circ}$ inclinations. Stokes $V$ contours are drawn at $\pm$0.1 and $\pm$1 mJy~beam$^{-1}$, and optical depth contours (over-plotted on the Stokes $I$ maps) are drawn at $\tau$ of 0.5, 0.75, and 1. In the bottom three panels, the major axis of the disk lies along the $\Delta x_{\rm obs}$-axis. Below each set of channel maps we include disk-integrated spectra. As described in the text, morphological differences between the vertical and toroidal field cases are readily apparent. Also of note, the edge-on toroidal case shows bright Stokes $V$ emission in the channel maps (with some regions producing $>10$ mJy/beam), but roughly zero signal in the integrated profile (due to spatial cancellation). This demonstrates the importance of leveraging spatial information when observing sources with sub-structured magnetic field configurations.}
  \label{fig:compChan}
\end{figure*}

The top two panels of Figure \ref{fig:compChan} show our results for face-on and intermediate inclination views of our purely vertical, $f_1 = 1$ and $f_2 = 0$, simulation.
In the face-on case, the Keplerian rotation of the disk is in the plane-of-the-sky, so its contribution to the line-of-sight velocity field is zero everywhere.
The emission is therefore spread in frequency space only due to line broadening, distributed  primarily among the central three channels.
Since $v_{\rm LOS, Kep} = 0$ km~s$^{-1}$ and the line-of-sight magnetic field is pointed entirely toward the observer at all locations, the $\boldsymbol{B}$-field configuration and viewing angle combination produces Stokes $I$ and $V$ profiles that are morphologically similar to the $\Delta \nu_z < \Delta \nu_{\rm line}$ case for the simple model (uniform magnetic field threading a uniform, non-moving parcel of gas) illustrated in Figure \ref{fig:splitDemo}.
Notably, in the central (zero velocity) channel the Stokes $V$ is zero due to $\sigma_+$ and $\sigma_-$ cancellation.

Unlike the face-on case, the intermediate ($i = 40^{\circ}$) inclination case produces line-of-sight velocity contributions.
For $\Delta x_{\rm obs} < 0$, $v_{\rm LOS, Kep} > 0$, and for $\Delta x_{\rm obs} > 0$, $v_{\rm LOS, Kep} < 0$, resulting in a double-peaked Stokes $I$ line profile.
Since the magnetic field here is again pointed in the same direction across all space (the inclination simply results in a $\cos{\theta_B}$ reduction of its line-of-sight strength), the shape of the Stokes $V$ profile is well mimicked by $dI/d\nu$.
Each channel in the Stokes $V$ map has positive and negative regions.
This pattern arises due to the varying amounts of red and blue shifted emission, and
can be understood most clearly by considering the central ($v_{\rm LOS} = 0$) channel.
In this channel, all the positive $V$ is located at $\Delta x_{\rm obs} < 0$ (where $v_{\rm LOS, Kep} > 0$) and all the negative $V$ is located at $\Delta x_{\rm obs} > 0$ (where $v_{\rm LOS, Kep} < 0$).
This flip occurs because, as demonstrated in Figure \ref{fig:splitDemo}, for a parcel of gas with line-of-sight velocity $v_0$, the peaks of the Stokes $V$ profile occur at $v_0 \pm \sim 0.4$ km/s (the precise value depends on the temperature and turbulence of the gas, which sets the slope of the Stokes $I$ over frequency).
As a result, the positive Stokes $V$ emission we observe in the zero velocity channel is dominated by red-shifted regions in the disk, and the negative Stokes $V$ arises in the blue-shifted regions.
In general, for a channel centered at $v = v_{\rm channel}$ the crossover ``line" from positive $V$ to negative $V$ occurs where $v_{\rm LOS} = v_{\rm channel}$.

\subsubsection{Toroidal Magnetic Field}\label{sssec:torCase}
In the bottom two panels of Figure \ref{fig:compChan}, we plot the results for our toroidal-only model ($f_1 = 0, f_2 = 0$).
Viewed edge-on ($i = 90^{\circ}$), we see the Stokes $V$ image is clearly split into four distinct regions in most channels.
The divide across the midplane (at $\Delta y = 0$) reflects the crossover from the magnetic field being oriented parallel to the Keplerian rotation to it being anti-parallel. Recall this feature aims to simulate $\boldsymbol{B}$-field ``wind-up" due to disk rotation.
Meanwhile, the vertical divide occurs because of the Keplerian rotation itself and is similar to the effect observed in the vertical magnetic field viewed at $i = 40^{\circ}$ case.
This divide is absent in the center-most channels, due to the co-locality of the velocity field and the magnetic field sign flips.
Together, these effects make it such that gas with slightly negative or slightly positive line-of-sight velocity components will both produce the same handedness of circular polarization at $v_{\rm LOS} = 0$.
Note also that the shape of the spatially integrated Stokes $V$ is no longer mimicked by $\frac{dI}{d\nu}$ due to the non-uniform magnetic field geometry.
We discuss this break down in more detail in Section \ref{sssec:fourB}.

Viewed at intermediate inclination, the emission from the toroidal $\boldsymbol{B}$-field is still split into four distinct sub-regions in most velocity channels.
This morphology arises because the CN slab traces out the disk surface, with each line of sight piercing the upper and lower surface at different radial positions, mirrored over 
the major axis of our axisymmetric disk.
When rotated to $i = 40^{\circ}$,
this arrangement gives four regions of coherent emission in the central velocity channel, because the magnetic field sign flips and velocity field sign flips are co-local (similar to the scenario for the \textit{two} regions for the edge-on case described above).
A few channels, e.g $\pm0.7$ km/s, express additional substructure.
This is due to the combined effects of the emitting layer height, the viewing geometry, and the Keplerian rotation.

\begin{figure*}
  \includegraphics[width=\textwidth]{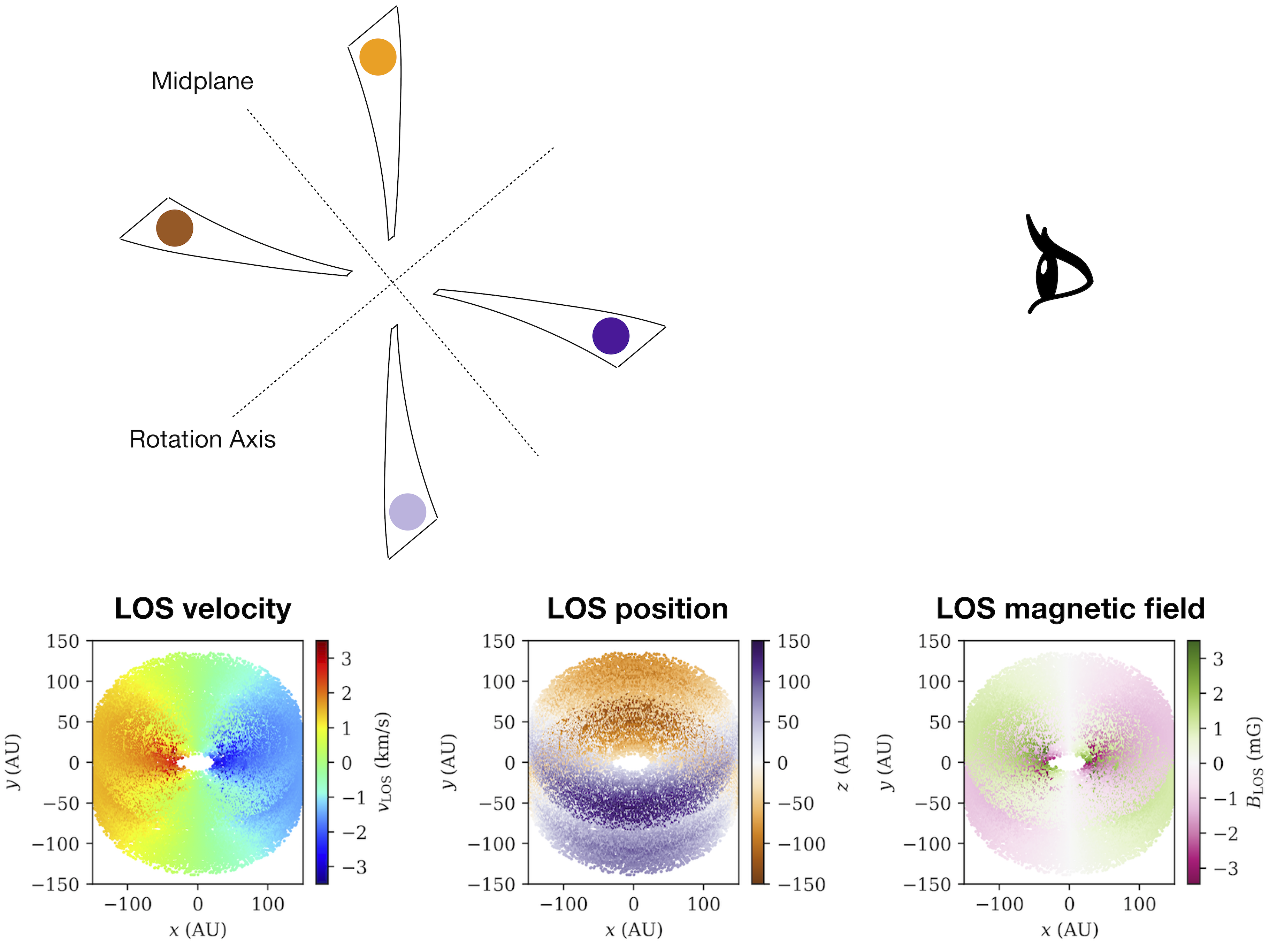}
  \caption{Illustration of the geometry of a disk with a toroidal magnetic field viewed at $i = 40^{\circ}$. \textit{Top:} Side view of the disk geometry. The four closed regions denote the locations where CN is placed in our model, with each one schematically color-coded to assist with interpretation of the ``LOS position" plot below. \textit{Bottom:} 3D Line-of-sight (LOS) velocity, LOS position, and LOS magnetic field maps. These visualizations are scatter plots, created by selecting 10,000 random locations in the disk, then color-coding the points accordingly and projecting them into the observer plane (notated as the $xy$-plane here). In the ``LOS position" plot, $z$ denotes LOS deviation from the center of the disk model space. The four CN slabs are clearly discernible, and this is why emission for the $i = 40^{\circ}$, toroidal case in Figure \ref{fig:compChan} is distributed into four distinct clumps (especially evident in the centermost channels). The clumps alternate between positive and negative $V$ because the magnetic field sign flips across the midplane, as illustrated in the ``LOS magnetic field" panel here.}
  \label{fig:torInter}
\end{figure*}

\subsection{Parameter Space Exploration}\label{ssec:parExp}
We now assess the observational impact of varying the parametric set-up of our model disk.
This analysis is performed in two parts.
First, we explore factors related to CN configuration and magnetic field strength (the first seven parameters listed in Table \ref{tab:parVary}).
Starting from our fiducial model (plotted in the top panel of Figure \ref{fig:fidModel}), which has a magnetic field component ratio of $B_{\rm vert}:B_{\rm tor}:B_{\rm rad} = 30\%:45\%:25\%$, we independently vary each parameter with the other parameters held fixed to examine \textit{par}ameter \textit{sli}ces (hereafter referred to as our \textit{parsli} analysis) through the model space.
This produces an easily digestible set of data to consider (as opposed to a full $n$-dimensional parameter space, it is instead $n$ 1-dimensional cuts).
In the subsequent section, we revert back to our fiducial model for those parameters and examine some different magnetic field geometries by varying $f_1$ and $f_2$.

\subsubsection{parsli}\label{sssec::parsli}
\begin{figure*}
  \makebox[\textwidth][c]{\includegraphics[width=1.2\textwidth]{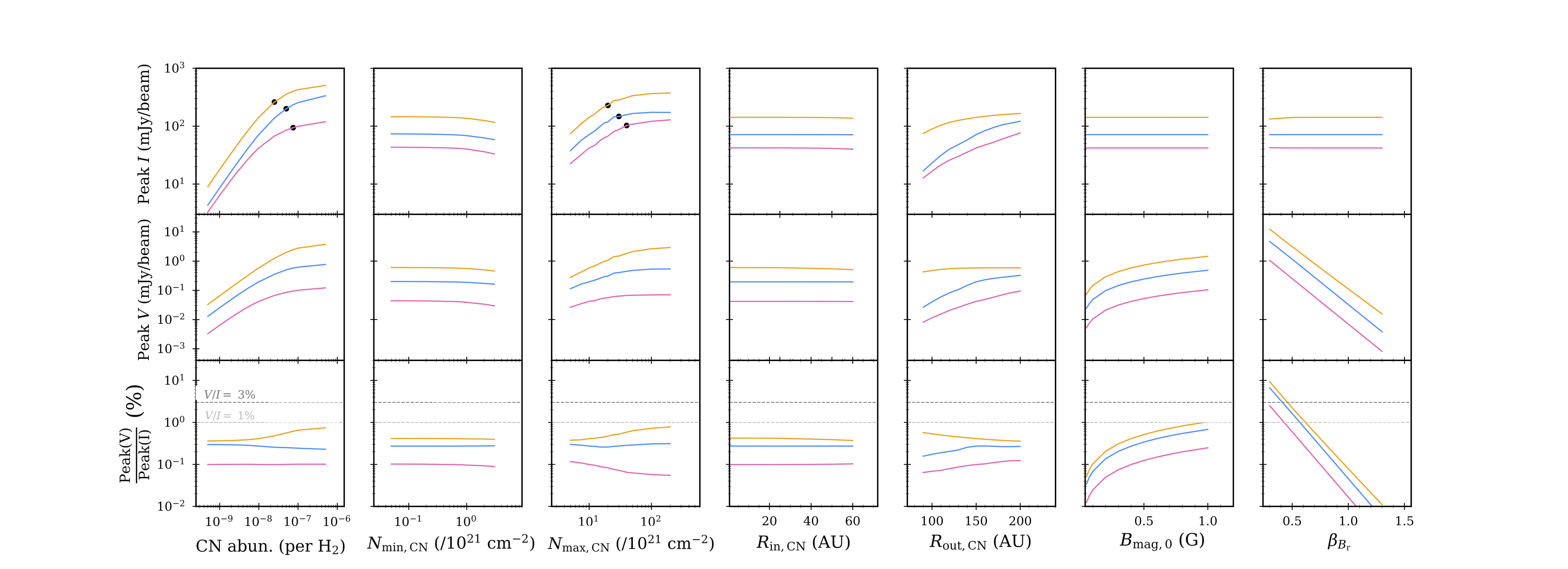}}
  \caption{Parameter space cuts for several variables, plotting the peak Stokes $I$, $V$, and $V$/$I$ for a 1'' beam as a function of parameter values. Peak flux is defined as the maximum value obtained for a given Zeeman simulation spatially and spectrally. The orange, blue, and pink curves correspond to $0^{\circ}$, $40^{\circ}$, and $90^{\circ}$ inclinations, respectively. In the top panel the black points indicate a transition to optically thick Stokes $I$. This only occurs for large X$_{\rm CN}$ or $N_{\rm max, CN}$.}
  \label{fig:parsli}
\end{figure*}

We vary the following parameters within the ranges specified in Table \ref{tab:parVary}: X$_{\rm CN}$, $R_{\rm in,CN}$, $R_{\rm out,CN}$, $N_{\rm min,CN}$, $N_{\rm max,CN}$, $B_{\rm sum, 0}$, and $\beta_{B_r}$.
After binning the simulation data to 0.4 km/s resolution and producing channel maps, we calculate the maximum flux (Stokes $I$ and $V$) and optical depth obtained for each model.
The results of these computations are provided in Figure \ref{fig:parsli}.
Among the parameters related to the distribution of CN in the disk, X$_{\rm CN}$, $R_{\rm out,CN}$, and $N_{\rm max,CN}$ are the most important.
Sensibly, flux scales linearly with CN abundance until there are enough molecules to produce $\tau \sim 1$, at which point optical depth effects become important and some of the emission is suppressed.
Extending the outer radius of the slab has a large effect due to the increase in emitting area. 
Extending the CN slab deeper into the disk to higher H$_2$ column densities (i.e., increasing $N_{\rm max,CN}$) incorporates more high density gas and thus also significantly boosts CN $J=1-0$ emission.

Meanwhile, there is very little dependence on the inner radius or the minimum hydrogen column density limits of the CN slab (i.e., the upper CN slab surface).
This is because there is a relatively small volume of gas at small $R$ (between the prescribed column density limits) and relatively low emissive material in the disk upper atmosphere.
Ultimately, for all these scaling relationships the operative quantity being modified is the total number of emitting CN molecules added or subtracted, so expansion of the CN slab into high density regions (or by a large volume) is what produces the largest increases in $I$ and, for a fixed magnetic field strength, $V$. Furthermore, we find that the magnetic quantities ($B_{\rm sum, 0}$ and $\beta_{B_r}$) scale with Stokes $V$ proportionally as expected from Eq.~\ref{eq:VB}.

\subsubsection{Extracting Magnetic Field Strengths}\label{sssec:fourB}
Given the complications of the disk magnetic structure, in this section we explore how the ``true'' value of the magnetic field put into the simulation compares to what one would extract using conventional line fitting techniques like Eq.~\ref{eq:VB}.
Included in this analysis are a subset\footnote{Some might say a sprig.} of the {\em parsli} simulations, including our purely toroidal (\texttt{tor}), purely vertical (\texttt{vert}), and fiducial (\texttt{fid}) models, as well as a ``fiducial-like" model with a boosted toroidal component (\texttt{fidtc}).

\begin{figure*}
  \centering
  \includegraphics[width=0.9\textwidth]{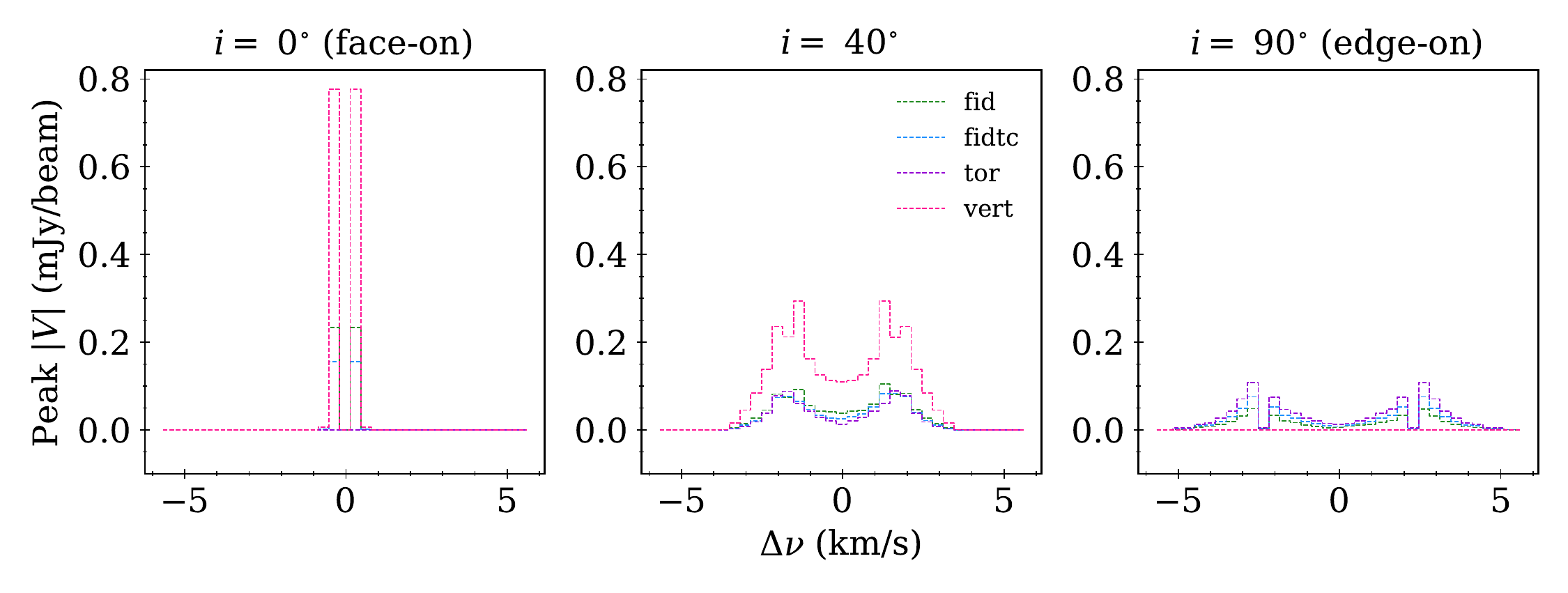}
  \caption{Peak $|V|$ as a function of frequency for each of the magnetic field configurations from Table \ref{tab:magConfigs}. The maximum value for each of these curves (i.e., the peak $|V|$ across all frequencies) is listed in Table \ref{tab:magConfData}. These data are binned to the same resolution (0.4 km/s) as the channel maps.}
  \label{fig:magConfigs}
\end{figure*}

All four configurations (summarized in Table \ref{tab:magConfigs}) have the same scaling for the magnetic field strength $B_{\rm sum,0} = 0.4$ Gauss and power law dependence $\beta_{B_r} = -0.75$, and therefore have the same mean (mass weighted) magnetic field strength of $B_{\rm avg} = 1.4$ mG within the CN emitting region.
Nevertheless, these models give different amounts of Stokes $V$ emission since the magnitude of the line-of-sight component of the magnetic field naturally changes.
In Table \ref{tab:magConfData} we list the mean line-of-sight magnetic field strength for each case, where
\begin{align}
    B_{\rm LOS, avg} =
        \frac{\int \frac{B_z + B_y \tan i}{\sqrt{1+\tan^2 i}}\rho(\boldsymbol{r}) d\boldsymbol{r}}{\int \rho(\boldsymbol{r}) d\boldsymbol{r}} \,,
\end{align}
integrated over the CN emitting region.
Due to symmetry, toroidal field components always produce $B_{\rm LOS, avg} = 0$.
Though this usefully expresses the importance of cancellation, most cancellation is due to spatial confusion rather than line-of-sight effects.
To get a sense of the magnitude of all the Zeeman-relevant emission, irrespective of whether $\boldsymbol{B}$ is directed toward or away from the observer, we also report the absolute value of the line-of-sight magnetic field strength as well, again integrated over the whole CN slab.
We also give the maximum Stokes $V$ found anywhere in the observation, with corresponding profiles (Peak $V$ vs. frequency) plotted in Figure \ref{fig:magConfigs}.

We find that the strongest Stokes $V$ emission in a given simulation is a strong function of the geometry of the underlying magnetic field.
Due to the lack of spatial cancellation within a given beam, vertical $\boldsymbol{B}$-field components produce by far the largest $B_{\rm LOS, avg}$ and peak $V$.
A face-on disk with a purely vertical magnetic yields a peak signal that is about a factor of seven larger than an edge-on disk with a purely toroidal magnetic field, even though $|B_{\rm LOS}|_{\rm, avg}$ is only $\approx$1.5 times larger.
In terms of producing a detection, intermediate inclination and edge-on viewing geometries are only preferable for field configurations that are almost entirely toroidal.
For our \texttt{fid} and \texttt{fidtc} cases, face-on observations result in emission that peaks 2.2 and 1.8 times higher than $i = 40^{\circ}$ observations, respectively.

Of course, due to Keplerian rotation, high inclination sources will have their emission distributed across a larger chunk of frequency space.
This geometry can be advantageous for some analysis goals, like localizing the emission along a given column of gas based on an assumed velocity profile \citep[e.g.,][]{teague2019}. However it can also have some disadvantages, like decreasing the line peak, thereby making detection more challenging.

\begin{figure*}
  \centering
  \includegraphics[width=0.9\textwidth]{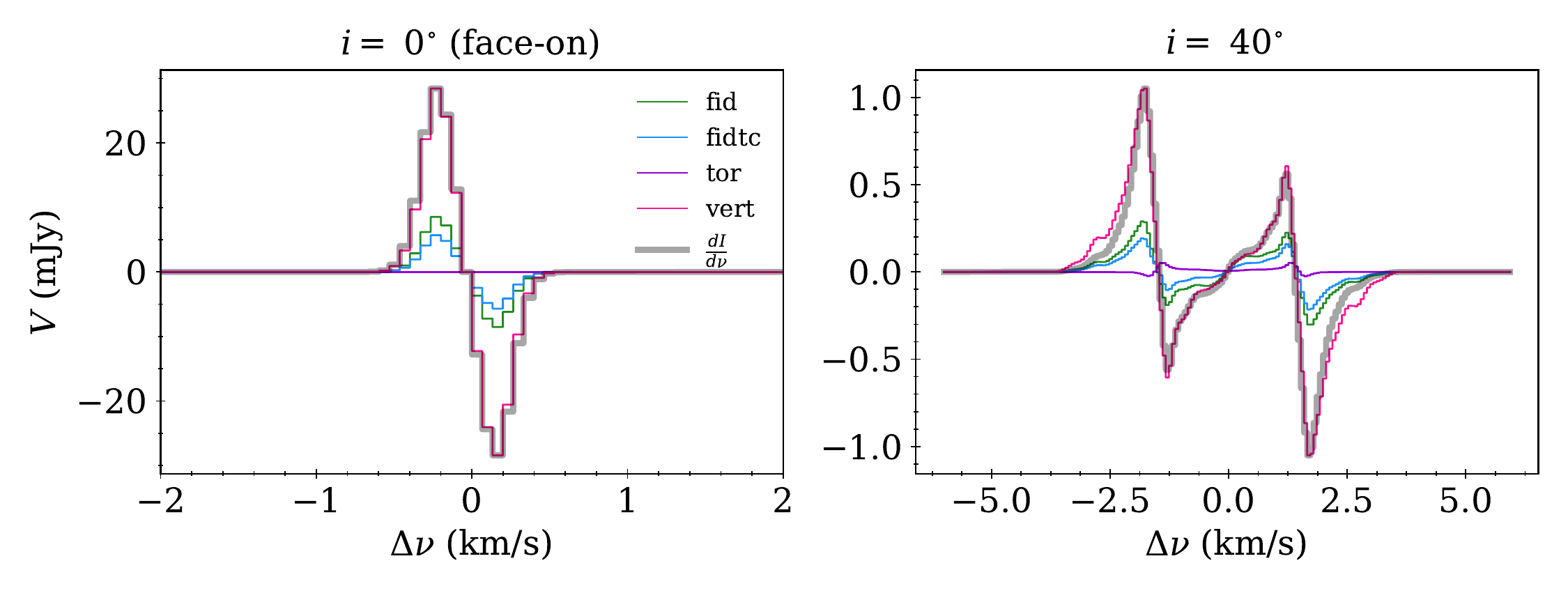}
  \caption{Spatially integrated Stokes $V$ profiles for the different magnetic field geometries we considered, as viewed at $i = 0^{\circ}$ and $i = 40^{\circ}$. Also plotted is the derivative of the Stokes $I$, scaled to fit the $V$ curve for a uniform magnetic field with a strength consistent with that put into our simulations. Magnetic field geometries with substructure produce significantly reduced Stokes $V$ magnitudes. If the toroidal field component is large enough, it can yield a profile that is different in shape from $dI/d\nu$.}
  \label{fig:fitB}
\end{figure*}

As described previously, fitting the derivative of the Stokes $I$ to the Stokes $V$ profile is a conventional technique for inferring line-of-sight magnetic field strengths from Zeeman observations (Eq~ \ref{eq:VB}).
This methodology may be applied to disk-scale observations, but we must be aware that the obtained $B_{\rm LOS}$ value may be significantly reduced due to field sub-structure in these environments.
In Figure \ref{fig:fitB} we plot the spatially integrated Stokes $V$ profiles for each of the magnetic field configurations, viewed at both face-on and intermediate inclinations. Note, the edge-on case produces $V \approx 0$~mJy for all four magnetic field geometries. This includes the toroidal field case due to the sign flip cancellation across the midplane.

In the same figure, we overplot the $dI/d\nu$ curve scaled to represent the $V$ inferred by setting $B = 1.4$ mG (the density-weighted average field strength for these runs).
In the face-on case, the shape of $dI/d\nu$ mimics the $V$ curves well because this view picks out the vertical field component, which is not subject to any cancellation.
The \texttt{fid} and \texttt{fidtc} curves are reduced in magnitude because they have a small fraction of their $\boldsymbol{B}$-field strength put into the vertical component.
At $i = 40^{\circ}$, the shape of $dI/d\nu$ still reasonably matches the \texttt{vert}, \texttt{fid}, \texttt{fidtc} Stokes $V$ profiles.
This highlights the dominance of the vertical field component, even when it is down to a factor 3.5 weaker than the toroidal component (as in the \texttt{fidtc} model).
However, in the fully toroidal model the profile is both substantially reduced and has a different morphology, owing to the sign flips in the magnetic field geometry.
If this magnetic sub-structure is not taken into account, fitting these curves using the conventional method results in considerable underestimates of the magnetic field strength.

\begin{deluxetable}{cccc}
\tablecaption{The magnetic field configurations we consider in Section \ref{sssec:fourB} and Section \ref{sssec:beam}. Percentages represent the fraction of the total magnetic field strength allocated to each of the components. \label{tab:magConfigs}}
\setlength{\tabcolsep}{5pt} 
\renewcommand{\arraystretch}{.9} 
\tablehead{\colhead{Name} & \colhead{Vertical (\%)} & \colhead{Toroidal (\%)} & \colhead{Radial (\%)}}
\startdata
 \texttt{vert}  & 100           & 0             & 0           \\
\texttt{tor}   & 0             & 100           & 0           \\
\texttt{fid}   & 30            & 45            & 25          \\
\texttt{fidtc} & 20            & 70            & 10          \\
\enddata
\end{deluxetable}

\begin{deluxetable}{ccccc}
\tablecaption{Mean line-of-sight magnetic field strength ($B_{\rm LOS, avg}$), mean absolute value of the line-of-sight magnetic field strength ($|B_{\rm LOS}|_{\rm, avg}$), and peak $|V|$ obtained from the channel maps for each of the magnetic field configurations we simulated. Values are reported for emission from the 113.144 GHz component only. A vertical field viewed face-on yields a peak $V$ flux that is a factor of $\sim7$ larger than a toroidal field viewed edge-on, even though $|B_{\rm LOS}|_{\rm, avg}$ is only a factor of $\sim1.6$ larger. This highlights the importance of cancellation for sub-structured (e.g., toroidal) magnetic field configurations. \label{tab:magConfData}}
\setlength{\tabcolsep}{5pt} 
\renewcommand{\arraystretch}{1} 
\tablehead{ &                                 & $i = 0^{\circ}$ & $i = 40^{\circ}$ & $i = 90^{\circ}$}
\startdata
      & $B_{\rm LOS, avg}$ (mG)         & 1.40            & 1.07             & 0                \\
\texttt{vert}  & $|B_{\rm LOS}|_{\rm, avg}$ (mG) & 1.40            & 1.07             & 0                \\
      & Peak $V$ (mJy/beam)             & 0.78            & 0.29             & 0                \\ \midrule
      & $B_{\rm LOS, avg}$ (mG)         & 0               & 0                & 0                \\
\texttt{tor}   & $|B_{\rm LOS}|_{\rm, avg}$ (mG) & 0               & 0.57             & 0.89                \\
      & Peak $V$ (mJy/beam)             & 0               & 0.09             & 0.11             \\ \midrule
      & $B_{\rm LOS, avg}$ (mG)         & 0.42            & 0.32             & 0                \\
\texttt{fid}   & $|B_{\rm LOS}|_{\rm, avg}$ (mG) & 0.42            & 0.37             & 0.46             \\
      & Peak $V$ (mJy/beam)             & 0.23            & 0.10             & 0.05             \\ \midrule
      & $B_{\rm LOS, avg}$ (mG)         & 0.28            & 0.21             & 0                \\
\texttt{fidtc} & $|B_{\rm LOS}|_{\rm, avg}$ (mG) & 0.28            & 0.43             & 0.63             \\
      & Peak $V$ (mJy/beam)             & 0.15            & 0.08             & 0.07             \\ \bottomrule
\enddata
\end{deluxetable}

\begin{figure*}
  \makebox[\textwidth][c]{\includegraphics[width=1.1\textwidth]{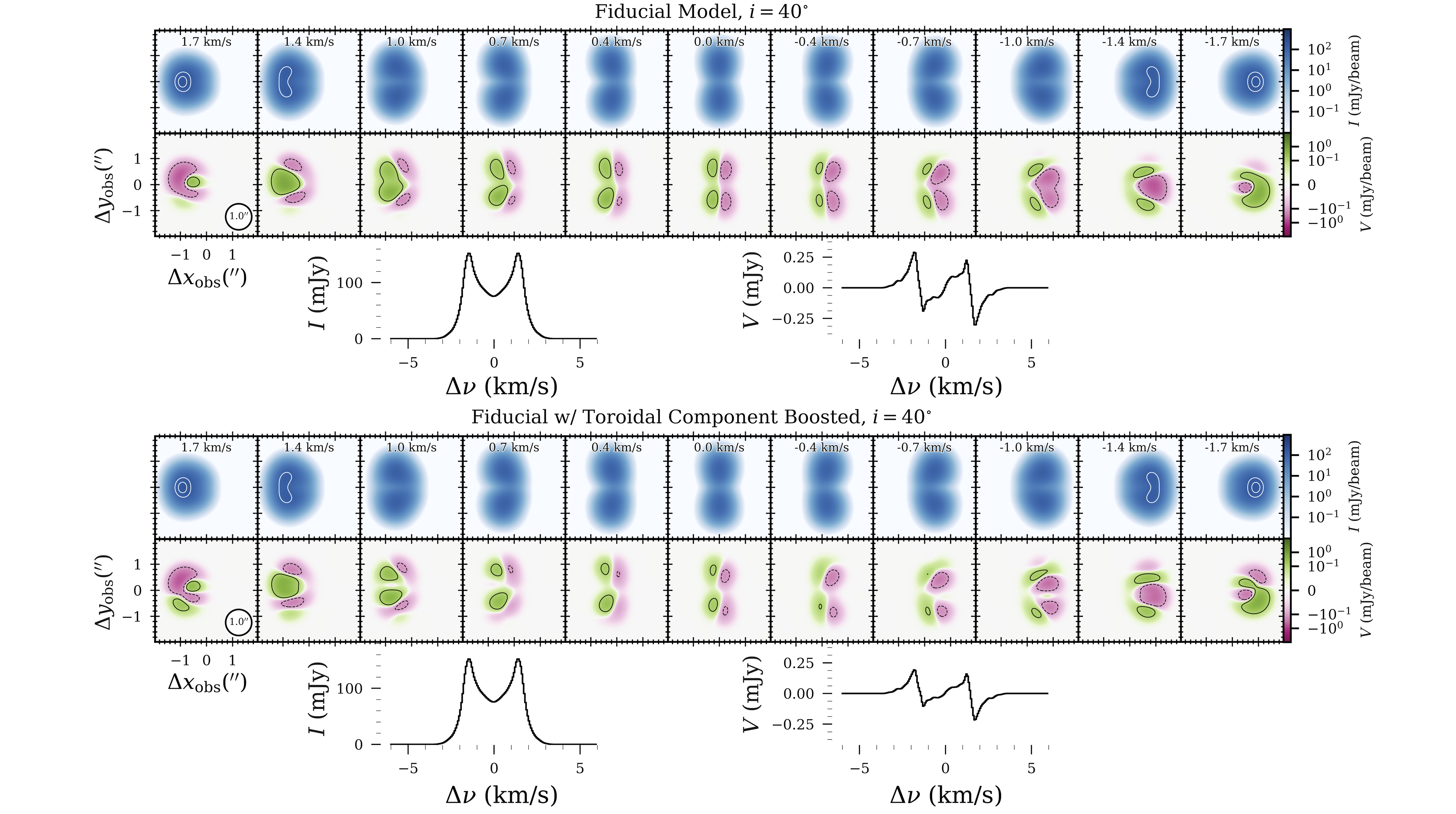}}
  \caption{\textit{Top panel:} Same as Figure \ref{fig:compChan}, now for our ``fiducial" case, viewed at $i = 40^{\circ}$. This model has its magnetic field strength divided such that $B_{\rm vert}:B_{\rm tor}:B_{\rm rad} = 30\%:45\%:25\%$. \textit{Bottom panel:} ``Toroidally boosted" version of the fiducial model, with $B_{\rm vert}:B_{\rm tor}:B_{\rm rad} = 20\%:70\%:10\%$.}
  \label{fig:fidModel}
\end{figure*}

\section{Discussion}\label{sec:discussion}

\subsection{Evidence of Magnetic Complexity in Stokes $V$ Channel Maps}\label{ssec:compDisc}
One of the principle results of this work is that channel map information from spatially resolved observations can be used to distinguish vertical and toroidal magnetic field geometries in intermediate inclination disks.
The features of the emission produced in these respective case studies are individually discussed in detail in Sections \ref{sssec:torCase} and \ref{sssec:vertCase}, but we can also use them to make a broader point about identifying magnetic sub-structure in general.
For the purely vertical $\boldsymbol{B}$-field geometry, each channel has exactly one region of positive $V$ and one region of negative $V$.
The spatial span of these regions changes for different channels (due to the Keplerian rotation of the disk), but at all velocities they are both continuous and symmetric about the major axis of the disk.
We can think of this as the ``unsubstructured" baseline --- that is, a rotating, axisymmetric disk with a uniform magnetic field threading through it will always produce Stokes $V$ channel map emission that looks like this.
Therefore, any deviation from this picture is suggestive of magnetic sub-structure.

The purely toroidal channel map is an extreme example of such deviation.
We see well-defined, interlaced regions of positive and negative $V$ emission, and the placement of these regions are not symmetric about the disk's major axis (due to the combined effects of CN positioning and viewing geometry, as illustrated in Figure \ref{fig:torInter}).
A key point here is that the morphology of the emission in the vertical $\boldsymbol{B}$-field case essentially only reflects the impact of Keplerian rotation (since the magnetic field is uniform), whereas the toroidal $\boldsymbol{B}$-field case is sensitive to the inherent nearside/farside asymmetries that arise in an inclined disk (since, unlike a uniform field, a sub-structured magnetic field is itself affected by the asymmetry).
Interestingly, this is also the reason why the toroidal $\boldsymbol{B}$-field case (at $i = 40^{\circ}$) does not have zero spatially-integrated $V$ emission.
Even though the disk's mean line-of-sight magnetic field strength is zero (see Table \ref{tab:magConfData}), the asymmetry results in non-zero emission for many velocity channels.

Our fiducial disk has a complicated magnetic field geometry ($B_{\rm vert}:B_{\rm tor}:B_{\rm rad} = 30\%:45\%:25\%$) and is intended to model a ``realistic" situation.
In the context of the discussion above, we can use it to make an important qualitative point about general interpretation of Zeeman observations in disk environments.
Looking at the channel map for the fiducial model (given in the top panel of Figure \ref{fig:fidModel}, as viewed at intermediate inclination), it is obvious that its morphology much more closely resembles the purely vertical case than the purely toroidal case.
This tells us that the observed Stokes $V$ will be dominated by any vertical field component, if present.
As a result, the shape of the integrated $V$ profile is almost identical to that of the purely vertical model.
However, as we know from the model set-up, the disk's intrinsic $\boldsymbol{B}$-field is \textit{not} primarily vertical --- only 30\% of the field strength is in the vertical component.
The only clear evidence of the other (sub-structured) components is the slight asymmetry in the Stokes $V$ emission across the disk's major axis.
This asymmetry is of course more pronounced if the toroidal component is boosted (as in the bottom panel of Figure \ref{fig:fidModel}), but even in that case the integrated $V$ profile shows virtually no evidence of the non-vertical magnetic field.
The channel map information therefore provides crucial context for interpreting $\boldsymbol{B}$-field orientation and strength.
It is important to be aware that even small asymmetries in the emission can represent a relatively high degree of complexity (and therefore cancellation) in the disk's intrinsic magnetic field.

\subsection{Detectability Analysis}\label{ssec:detect}
Apart from the characteristics of the source itself, there are a few observational effects that can play a role in governing the level of detectability for our emission of interest.
We first evaluate the importance of beam size, then discuss the potential efficacy of velocity-based stacking of the hyperfine transitions (listed in Table \ref{tab:CNLines}) to boost the total Stokes $V$ flux.

\subsubsection{Beam Size}\label{sssec:beam}
In the case of observations for which the total emission is the quantity of interest, there is a direct proportionality between the size of the beam, $\theta_{\rm beam}$, and the maximum flux observed per beam.
This relationship is not necessarily true for observations of the Stokes $V$, because the positive and negative components of the emission become more prone to cancellation when integrated over more area.
Therefore, larger beams are liable to wash out signals of opposite polarity.

\begin{figure*}
  \includegraphics[width=\textwidth]{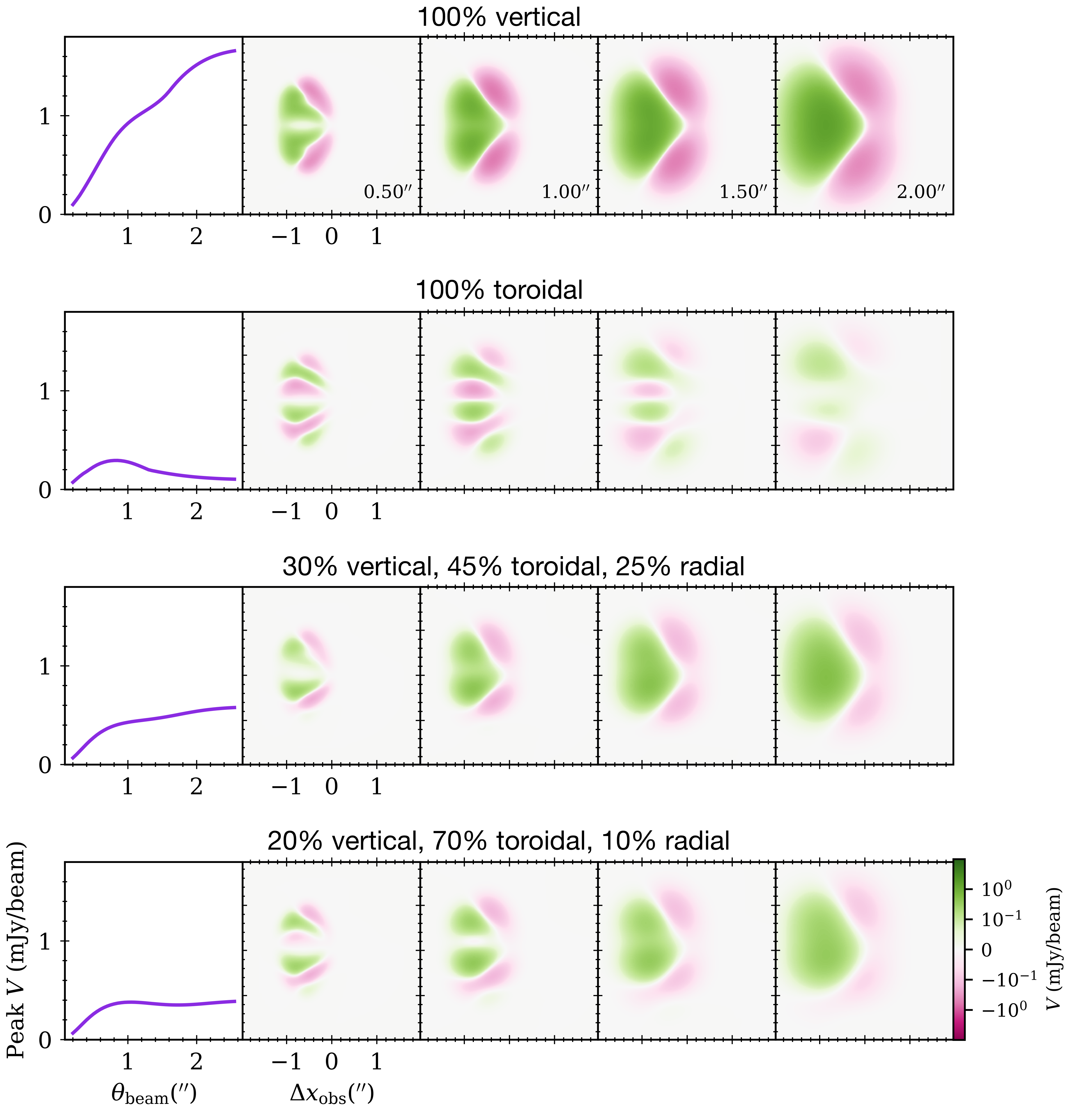}
  \caption{Comparison plots of a 0.4 km/s wide channel (centered at 1 km/s) for several choices beam size, viewed at intermediate ($40^{\circ}$) inclination. Each row reflects a different magnetic field geometry. The left panel shows how the maximum observable intensity (e.g., flux coming from the brightest pixel) changes as a function of beam size. Note that in the fully toroidal case, there is a turnover in Peak $V$ at $\theta_{\rm beam} = 0.8$ arcsec. This demonstrates the importance of spatial cancellation in poorly resolved observations of sources with toroidally dominated magnetic fields.}
  \label{fig:beamComp}
\end{figure*}

In Figure \ref{fig:beamComp}, we choose a representative velocity channel (0.4 km/s wide, centered at 1 km/s) and for each of the magnetic field configurations discussed in Section \ref{sssec:fourB} show $V$ emission maps using $\theta_{\rm beam} = 0.5,1,1.5$ and $2''$, viewed at $i = 40^{\circ}$.
We also plot peak $V$ (mJy/beam) vs. $\theta_{\rm beam}$.
In the 100\% vertical magnetic field simulation, the emission scales approximately linearly with the size of the beam.
This scaling occurs because the magnetic field has uniform direction in this case, and as such there is no sub-structure to produce cancellation.
In the other models we introduce toroidal (and radial) $\boldsymbol{B}$-field components, and the impact this has in suppressing signal is clear.
The most striking example is the 100\% toroidal case, for which we observe a turnover in peak $V$ at $\theta_{\rm beam} \approx 0.8''$.
The signal becomes almost completely washed out for very large beams.
For the more complicated magnetic field geometries, the $V$ vs. $\theta_{\rm beam}$ plots for those cases exhibit a knee at $\approx 0.8''$, the scale at which toroidal field cancellation becomes important.
For larger $\theta_{\rm beam}$, the rate of increase of the $V$ emission tapers considerably.
Since simulations generally predict substantial toroidal $\boldsymbol{B}$-field components, these results suggest that $\theta_{\rm beam} \approx 0.8''$ is the most reasonable choice for observations to maximize signal and preserve good spatial resolution when little is known about the true magnetic field geometry. 

\subsubsection{Hyperfine Component Stacking}\label{sssec:transitions}
\begin{figure*}
  \includegraphics[width=\textwidth]{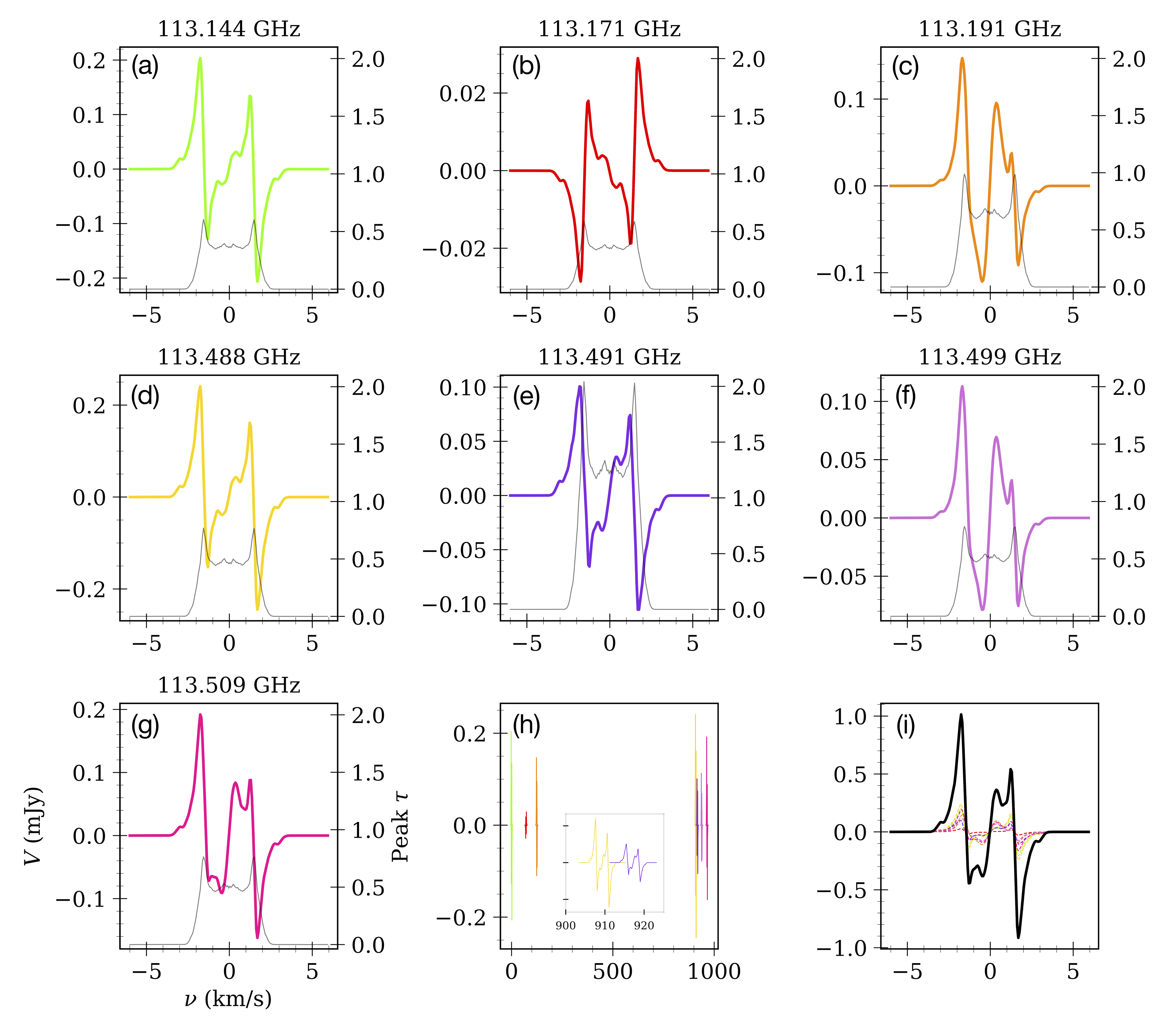}
  \caption{\textbf{Panels (a) - (g):} Stokes $V$ profiles for each of the CN $J = 1 - 0$ hyperfine transitions. Also included in each panel are optical depth profiles, plotting the peak $\tau$ (across all space) found at each frequency. \textbf{Panel (h):} A plot of where the lines lie in frequency space with respect to each other. They are mostly well separated. In the sub-panel we show that the 113.488 GHz and 113.491 GHz, which are relatively nearby, are still completely non-blended. \textbf{Panel (i):} Stacked profile of all 7 lines. Note that because the 113.171 GHz transition has negative $z_B$, its profile should be negated before stacking.}
  \label{fig:trans}
\end{figure*}

For the CN $J = 1-0$ transition, there are 7 observable hyperfine components.
So far we have only considered the 113.144 GHz line (as a representative case), but it is in principle possible to leverage the flux from multiple lines to produce a stronger detection.
In Figure \ref{fig:trans} we plot the (spatially integrated) line flux results of simulations for all the transitions, performed for our fiducial model at $i = 40^{\circ}$.
Since the lines are entirely non-blended, stacking is possible.
The stacked line profile has a peak flux that is a factor of $\sim$5 larger than that produced solely by the 113.144 GHz component.

\subsection{Sub-structured Gas Distribution}\label{ssec:gas}
\begin{figure*}
  \includegraphics[width=\textwidth]{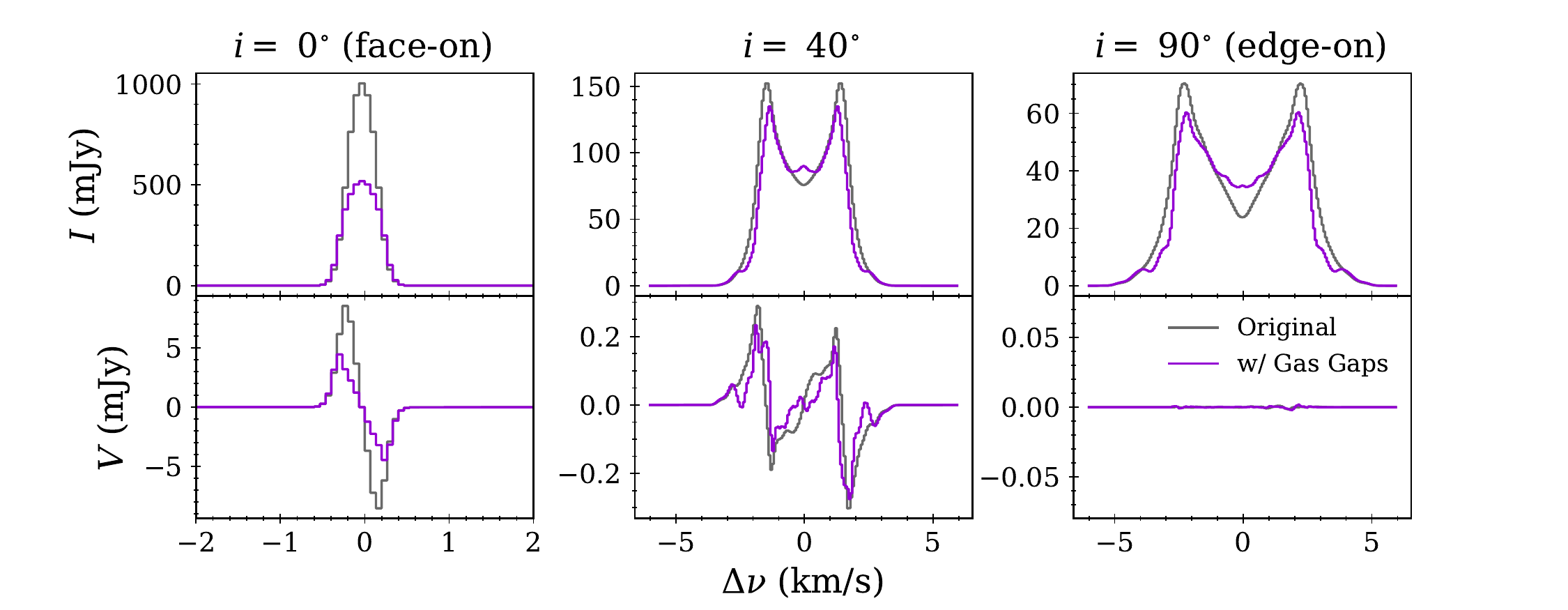}
  \caption{Comparison plots of the spatially integrated Stokes $I$ and $V$ profiles for a model with smooth gas distributions (e.g., \texttt{fid}) and one with gas gaps added. Due to increased opacity in the ring regions, the face-on view yields significantly reduced emission when gas sub-structure is introduced. This effect is present in the intermediate and edge-on cases as well, but to a smaller extent since the emission is spread over a larger range of velocity space. At $i = 40^{\circ}$, the gas gaps affect the morphology of the $V$ profile as well.}
  \label{fig:gasGaps}
\end{figure*}

Our fiducial disk includes rings in the large dust population.
As part of our modeling work we also tested disk scenarios with smooth (re-normalized to the same mass) large dust distributions, and found that the presence or absence of dust sub-structure has a negligible effect on the line emission results.
However, it is possible that this sub-structure may exist in the gas as well.
Observations of C$^{18}$O ($ J = 1-0$) emission in AS 209 by \citet{favre2019} show evidence of gas deficits that are spatially coincident with the dust gaps.
To model this scenario, we ran additional versions of our fiducial simulations with gas density gaps carved out according to the $\delta(R)$ prescription given in Equation \ref{eq:delta}.
The density distribution is renormalized such that the total gas mass is kept the same as it was in the original runs.
In Figure \ref{fig:gasGaps} we compare the emission profiles from these sub-structured runs with the original smooth ones.

For intermediate inclination and edge-on models, the Stokes $I$ is redistributed in velocity space when sub-structure is introduced, yielding more ``peaky" profiles since more of the CN gas is constrained to specific radii.
The opacity in these regions is slightly higher, exceeding $\tau = 1$ only near the peaks (this results in a $\sim20$\% lower maximum in $I$ than the original).
For most frequencies the emission remains optically thin, but there are still differences in the profile morphology as a result of the added gas sub-structure.
This is an important point to consider --- in the case of sub-structured disks, it is possible that some of the features in the $V$ profile are \textit{not} the result of magnetic complexity.
Observers should be cautious of this when searching for signatures of $\boldsymbol{B}$-field morphology in their data.

For the face-on model, the opacity increases dramatically with the addition of gaps.
This is because the emission, already distributed over a relatively narrow range in frequency space (since $v_{\rm LOS} = 0$ everywhere), is now pushed to smaller regions in observer space.
As a result of these optical depth effects, the $I$ and $V$ emission are both reduced significantly (by a factor of $\sim2$).

\subsection{Comparison to ALMA Percentage Polarization Limits}\label{ssec:alma}
ALMA's current circular polarization instrumentation is nominally stated to have a 1.8\% percentage polarization limit.
In the bottom row of Figure \ref{fig:parsli}, we give percentage polarization for the models in our main \textit{parsli} grid.
Since values for $I$ and $V$ vary across the observer plane, we report peak values for each run.
Our fiducial model yields percentages of 0.1\%, 0.3\%, and 0.4\% for 90$^{\circ}$ (edge-on), 40$^{\circ}$, and 0$^{\circ}$ (face-on) viewing angles, respectively.

Increasing the CN abundance or the depth of the CN slab (to larger $N_{\rm max,CN}$) increases $V/I$ in the face-on case, and extending the maximum radius of the slab leads to larger $V/I$ in the intermediate inclination and edge-on cases.
Peak percentage polarization also scales with $B_{\rm sum,0}$, of course.
Increasing the values of these parameters in various combinations produces a parameter space of optimistic disk scenarios that reach the nominal ALMA limit of 1.8\%.
For instance, if we set $B_{\rm sum,0}$ to 1.0 G (corresponding to $B_{\rm avg} = 3.5$ mG), we could produce 1.8\% polarization by also increasing the CN abundance to $\approx 3 \times 10^{-7}$ (per H$_{2}$) or increasing $N_{\rm max,CN}$ to about $10^{23}$ cm$^{-2}$.
It should be noted that at these high values of CN abundance and $N_{\rm max,CN}$, opacity effects will start to come into play as some regions of the disk reach $\tau > 1$.

Based on their circular polarization (non-detection) observations of TW Hydra,  \citet{vlemmings2019} suggest ALMA may be capable of substantially better polarization performance, inferring a $<$0.8\% detection level.
For our face-on fiducial model, 0.8\% polarization can be reached if we set $B_{\rm sum,0} = 0.8$ G, which corresponds to a mean magnetic field in the CN emitting region of $B_{\rm avg} = 2.8$ mG.
This agrees reasonably well with the 2.6 mG limit \citet{vlemmings2019} report.
We note however that, as discussed above, there are also factors related to the disk set-up that can affect percentage polarization --- namely the abundance of the emitting molecule and the depth of the molecular layer.

\section{Conclusions}\label{sec:conclusions}
We simulated the Stokes $I$ and $V$ CN $J = 1-0$ emission arising from a ringed disk (modelled after the AS 209 disk system) with the POLARIS radiative transfer code. We produced synthetic observations viewed at face-on, intermediate ($i = 40^{\circ}$), and edge-on inclinations.
We varied several parameters in our model to probe how the emission changes as a function of the magnetic field configuration and the properties of the CN emitting region.
Our main conclusions are as follows:
\begin{enumerate}
    \item Vertical and toroidal magnetic field configurations produce substantially different Stokes $V$ emission, and it is possible to distinguish them based on channel map morphology. At intermediate inclination, vertical $\boldsymbol{B}$-field components produce blotches of positive and negative $V$ emission that are symmetric about the major axis of the disk. Asymmetries to this end are a telltale sign of magnetic complexity, and even small ones can signify a relatively strong toroidal magnetic field component.
    For sources with both vertical and toroidal components, the toroidal component must be much stronger than the vertical component for it to contribute significantly to the spatially integrated Stokes $V$ emission, unless the disk is viewed close to edge-on.
    \item For our fiducial disk model, which has ``realistic" distributions of magnetic field strength and CN, the maximum Stokes $V$ signal obtained from our synthetic observations (at 0.4 km/s velocity resolution, with a $1''$ beam) is 0.6, 0.2, and 0.04 mJy/beam for face-on, $i=40^{\circ}$, and edge-on observations, respectively. Note that these values are for the 113.144 GHz transition only --- considering the other hyperfine components can fruitfully improve the signal (see item 6  below).
    \item The Stokes $V$ scales with the strength of the magnetic field, and both the Stokes $I$ and Stokes $V$ scale with the total number of CN molecules. For our fiducial model the line emission is optically thin, but if CN exists deep enough into the disk (at column densities $\gtrsim 3 \times 10^{22}$ cm$^{-2}$) or if it is abundant enough ($\gtrsim 4 \times 10^{-8}$ CN molecules per H$_2$), the emission can transition to optically thick in some regions.
    \item The traditional method for inferring magnetic field strength from Zeeman observations (i.e., fitting with Equation \ref{eq:VB}) must be approached with caution in disk environments, because PPDs are expected to have significant magnetic sub-structure. If the magnetic field has a strong vertical component, this component will be picked out effectively for face-on or intermediate inclination observations.  However, its magnitude will imply a magnetic field strength that may be significantly reduced from the true value, depending on how much of the field is distributed into the other components. For close to edge-on sources or disks with dominant toroidal fields, the spatially integrated Stokes $V$ profile will be greatly diminished due to cancellation, and its shape will not be matched by $dI/d\nu$ due to the non-uniformity of the magnetic field. In this case, leveraging spatial information becomes crucial.
    \item Choice of beam size can play an important role in the detectability of the Stokes $V$ emission in sources with magnetic sub-structure. If the magnetic field is toroidally dominated, there is a turnover in flux per beam at $\theta_{\rm beam} \approx 0.8''$ in our model. This beam size corresponds to a physical size of $\sim100$ au. Larger beams wash out the signal due to cancellation.
    \item The 7 observable hyperfine components in the CN $J = 1 - 0$ suite are well-resolved in frequency space.
    Due to optical depth effects and differing critical densities, the profiles of these components are not all identical. Nonetheless, they are similar enough that stacking is feasible. We demonstrate that stacking can increase the total signal by a factor of $\sim$5 over just using the strongest 113.144 GHz line.
    \item The presence of gas sub-structure in the disk can have important effects on the Stokes $V$ emission, both in terms of magnitude and morphology. Face-on disks with gaps have substantially elevated optical depth (in the rings) compared to equal mass gap-less counterparts. If some regions (i.e., the rings) reach $\tau > 1$, this is liable to produce reduced emission in the spatially integrated profile. Intermediate inclination disks are also susceptible to this effect, but to a lesser extent since the emission is spread over a wider breadth of frequency space. As our $i = 40^{\circ}$ simulation shows, gas gaps in intermediate inclination sources also produce perturbations in the Stokes $V$ profile, which could in principle be interpreted (incorrectly) as evidence of magnetic sub-structure. Observers should be cautioned of this when inferring magnetic field information from Zeeman observations.
\end{enumerate}

In this work we considered one disk structure and only performed line emission simulations.  Natural future extensions could include testing different density distributions (in both gas and dust) and simulating the continuum emission.
Namely, one potentially important factor we have not accounted for here is that some sources may have thick dust midplanes that could block up to half of the disk, depending on the viewing geometry.
This could of course reduce total emission, but also may eliminate some of the cancellation that occurs in the Stokes $V$ emission of sub-structured magnetic field configurations, which could have interesting effects on both the morphology and detectability of the signal.
In the simulations we performed for this work the midplane was optically thin at 113 GHz, so dust did not play a role in the radiative transfer beyond factoring into the calculation of the dust and gas temperature. However, future simulations of Zeeman at higher $J$ rotational transitions should take the possibility of optically thick continuum emission into account when simulating line observations.

\section*{Acknowledgements}
RRM is supported in part by the National Radio Astronomy Observatory (NRAO) through a Student Observing Support (SOS) award and by NASA 80NSSC18K0481. RRM acknowledges additional support from the Virginia Space Grant Consortium (VSGC) Graduate Fellowship Award, as well as useful communication with Crystal Brogan, Scott Suriano, Daniel Lin, Robert Brauer, and Stefan Reissl. LIC gratefully acknowledges support from the David and Lucille Packard Foundation and the VSGC New Investigators Award and NASA 80NSSC20K0529. ZYL is supported in part by NASA 80NSSC20K0533 and NSF AST-1815784.

\bibliographystyle{aasjournal}

\end{document}